\documentclass[journal,draftclsnofoot,onecolumn,12pt]{IEEEtran}

\usepackage{graphicx}
\usepackage{subfigure}
\usepackage{multirow}
\usepackage{array}
\newcolumntype{P}[1]{>{\centering\arraybackslash}p{#1}}
\newcolumntype{M}[1]{>{\centering\arraybackslash}m{#1}}

\usepackage{amsthm,amssymb,amsmath,graphicx,multirow,color,amsfonts}%
\usepackage[update,prepend]{epstopdf}
\usepackage{multirow}
\usepackage[latin1]{inputenc}
\usepackage{tikz}
\usepackage{bbm} 
\usepackage{pdfpages}
\usepackage{multirow}
\usepackage{subfig}
\usepackage{comment}

\usepackage{graphicx}
\usepackage{multicol}
\usepackage{cite}

\usepackage[justification=centering]{caption}
\usepackage{textcomp}
\usepackage{psfrag}
\usepackage{arydshln}
\usepackage{url}
\usepackage{soul}
\usepackage{graphicx,color}
\usepackage[nolist]{acronym}
\usepackage{algorithm,algorithmic} 

\usepackage{mathtools,lipsum}
\usepackage{cuted}
\usepackage{amsmath}
\newtheorem{proposition}{Proposition} 
\usepackage{mathrsfs}

\usepackage[capitalise]{cleveref}
\Crefname{equation}{Eq.\!}{Eqs.\!}
\Crefname{figure}{Fig.\!}{Figs.\!}
\Crefname{tabular}{Tab.\!}{Tabs.\!}
\Crefname{section}{Section\!}{Sections.\!}

\def\nb0{{\mathbf{0}}}
\def\nb1{{\mathbf{1}}}

\newtheorem{lemma}{Lemma}

\newtheorem{definition}{Definition}

\newtheorem{theorem}{Theorem}

\newenvironment{sequation}{
\begin{equation}\small}{\end{equation}
}

\begin{document}
\graphicspath{{./Figures/}}
	\begin{acronym}

\acro{5G-NR}{5G New Radio}
\acro{3GPP}{3rd Generation Partnership Project}
\acro{ABS}{aerial base station}
\acro{AC}{address coding}
\acro{ACF}{autocorrelation function}
\acro{ACR}{autocorrelation receiver}
\acro{ADC}{analog-to-digital converter}
\acrodef{aic}[AIC]{Analog-to-Information Converter}     
\acro{AIC}[AIC]{Akaike information criterion}
\acro{aric}[ARIC]{asymmetric restricted isometry constant}
\acro{arip}[ARIP]{asymmetric restricted isometry property}

\acro{ARQ}{Automatic Repeat Request}
\acro{AUB}{asymptotic union bound}
\acrodef{awgn}[AWGN]{Additive White Gaussian Noise}     
\acro{AWGN}{additive white Gaussian noise}

\acro{APSK}[PSK]{asymmetric PSK} 

\acro{waric}[AWRICs]{asymmetric weak restricted isometry constants}
\acro{warip}[AWRIP]{asymmetric weak restricted isometry property}
\acro{BCH}{Bose, Chaudhuri, and Hocquenghem}        
\acro{BCHC}[BCHSC]{BCH based source coding}
\acro{BEP}{bit error probability}
\acro{BFC}{block fading channel}
\acro{BG}[BG]{Bernoulli-Gaussian}
\acro{BGG}{Bernoulli-Generalized Gaussian}
\acro{BPAM}{binary pulse amplitude modulation}
\acro{BPDN}{Basis Pursuit Denoising}
\acro{BPPM}{binary pulse position modulation}
\acro{BPSK}{Binary Phase Shift Keying}
\acro{BPZF}{bandpass zonal filter}
\acro{BSC}{binary symmetric channels}              
\acro{BU}[BU]{Bernoulli-uniform}
\acro{BER}{bit error rate}
\acro{BS}{base station}
\acro{BW}{BandWidth}
\acro{BLLL}{ binary log-linear learning }

\acro{CP}{Cyclic Prefix}
\acrodef{cdf}[CDF]{cumulative distribution function}   
\acro{CDF}{Cumulative Distribution Function}
\acrodef{c.d.f.}[CDF]{cumulative distribution function}
\acro{CCDF}{complementary cumulative distribution function}
\acrodef{ccdf}[CCDF]{complementary CDF}               
\acrodef{c.c.d.f.}[CCDF]{complementary cumulative distribution function}
\acro{CD}{cooperative diversity}

\acro{CDMA}{Code Division Multiple Access}
\acro{ch.f.}{characteristic function}
\acro{CIR}{channel impulse response}
\acro{cosamp}[CoSaMP]{compressive sampling matching pursuit}
\acro{CR}{cognitive radio}
\acro{cs}[CS]{compressed sensing}                   
\acrodef{cscapital}[CS]{Compressed sensing} 
\acrodef{CS}[CS]{compressed sensing}
\acro{CSI}{channel state information}
\acro{CCSDS}{consultative committee for space data systems}
\acro{CC}{convolutional coding}
\acro{Covid19}[COVID-19]{Coronavirus disease}

\acro{DAA}{detect and avoid}
\acro{DAB}{digital audio broadcasting}
\acro{DCT}{discrete cosine transform}
\acro{dft}[DFT]{discrete Fourier transform}
\acro{DR}{distortion-rate}
\acro{DS}{direct sequence}
\acro{DS-SS}{direct-sequence spread-spectrum}
\acro{DTR}{differential transmitted-reference}
\acro{DVB-H}{digital video broadcasting\,--\,handheld}
\acro{DVB-T}{digital video broadcasting\,--\,terrestrial}
\acro{DL}{DownLink}
\acro{DSSS}{Direct Sequence Spread Spectrum}
\acro{DFT-s-OFDM}{Discrete Fourier Transform-spread-Orthogonal Frequency Division Multiplexing}
\acro{DAS}{Distributed Antenna System}
\acro{DNA}{DeoxyriboNucleic Acid}

\acro{EC}{European Commission}
\acro{EED}[EED]{exact eigenvalues distribution}
\acro{EIRP}{Equivalent Isotropically Radiated Power}
\acro{ELP}{equivalent low-pass}
\acro{eMBB}{Enhanced Mobile Broadband}
\acro{EMF}{ElectroMagnetic Field}
\acro{EU}{European union}
\acro{EI}{Exposure Index}
\acro{eICIC}{enhanced Inter-Cell Interference Coordination}

\acro{FC}[FC]{fusion center}
\acro{FCC}{Federal Communications Commission}
\acro{FEC}{forward error correction}
\acro{FFT}{fast Fourier transform}
\acro{FH}{frequency-hopping}
\acro{FH-SS}{frequency-hopping spread-spectrum}
\acrodef{FS}{Frame synchronization}
\acro{FSsmall}[FS]{frame synchronization}  
\acro{FDMA}{Frequency Division Multiple Access}

\acro{GA}{Gaussian approximation}
\acro{GF}{Galois field }
\acro{GG}{Generalized-Gaussian}
\acro{GIC}[GIC]{generalized information criterion}
\acro{GLRT}{generalized likelihood ratio test}
\acro{GPS}{Global Positioning System}
\acro{GMSK}{Gaussian Minimum Shift Keying}
\acro{GSMA}{Global System for Mobile communications Association}
\acro{GS}{ground station}
\acro{GMG}{ Grid-connected MicroGeneration}

\acro{HAP}{high altitude platform}
\acro{HetNet}{Heterogeneous network}

\acro{IDR}{information distortion-rate}
\acro{IFFT}{inverse fast Fourier transform}
\acro{iht}[IHT]{iterative hard thresholding}
\acro{i.i.d.}{independent, identically distributed}
\acro{IoT}{Internet of Things}                      
\acro{IR}{impulse radio}
\acro{lric}[LRIC]{lower restricted isometry constant}
\acro{lrict}[LRICt]{lower restricted isometry constant threshold}
\acro{ISI}{intersymbol interference}
\acro{ITU}{International Telecommunication Union}
\acro{ICNIRP}{International Commission on Non-Ionizing Radiation Protection}
\acro{IEEE}{Institute of Electrical and Electronics Engineers}
\acro{ICES}{IEEE international committee on electromagnetic safety}
\acro{IEC}{International Electrotechnical Commission}
\acro{IARC}{International Agency on Research on Cancer}
\acro{IS-95}{Interim Standard 95}

\acro{KPI}{Key Performance Indicator}

\acro{LEO}{low earth orbit}
\acro{LF}{likelihood function}
\acro{LLF}{log-likelihood function}
\acro{LLR}{log-likelihood ratio}
\acro{LLRT}{log-likelihood ratio test}
\acro{LoS}{Line-of-Sight}
\acro{LRT}{likelihood ratio test}
\acro{wlric}[LWRIC]{lower weak restricted isometry constant}
\acro{wlrict}[LWRICt]{LWRIC threshold}
\acro{LPWAN}{Low Power Wide Area Network}
\acro{LoRaWAN}{Low power long Range Wide Area Network}
\acro{NLoS}{Non-Line-of-Sight}
\acro{LiFi}[Li-Fi]{light-fidelity}
 \acro{LED}{light emitting diode}
 \acro{LABS}{LoS transmission with each ABS}
 \acro{NLABS}{NLoS transmission with each ABS}

\acro{MB}{multiband}
\acro{MC}{macro cell}
\acro{MDS}{mixed distributed source}
\acro{MF}{matched filter}
\acro{m.g.f.}{moment generating function}
\acro{MI}{mutual information}
\acro{MIMO}{Multiple-Input Multiple-Output}
\acro{MISO}{multiple-input single-output}
\acrodef{maxs}[MJSO]{maximum joint support cardinality}                       
\acro{ML}[ML]{maximum likelihood}
\acro{MMSE}{minimum mean-square error}
\acro{MMV}{multiple measurement vectors}
\acrodef{MOS}{model order selection}
\acro{M-PSK}[${M}$-PSK]{$M$-ary phase shift keying}                       
\acro{M-APSK}[${M}$-PSK]{$M$-ary asymmetric PSK} 
\acro{MP}{ multi-period}
\acro{MINLP}{mixed integer non-linear programming}

\acro{M-QAM}[$M$-QAM]{$M$-ary quadrature amplitude modulation}
\acro{MRC}{maximal ratio combiner}                  
\acro{maxs}[MSO]{maximum sparsity order}                                      
\acro{M2M}{Machine-to-Machine}                                                
\acro{MUI}{multi-user interference}
\acro{mMTC}{massive Machine Type Communications}      
\acro{mm-Wave}{millimeter-wave}
\acro{MP}{mobile phone}
\acro{MPE}{maximum permissible exposure}
\acro{MAC}{media access control}
\acro{NB}{narrowband}
\acro{NBI}{narrowband interference}
\acro{NLA}{nonlinear sparse approximation}
\acro{NLOS}{Non-Line of Sight}
\acro{NTIA}{National Telecommunications and Information Administration}
\acro{NTP}{National Toxicology Program}
\acro{NHS}{National Health Service}

\acro{LOS}{Line of Sight}

\acro{OC}{optimum combining}                             
\acro{OC}{optimum combining}
\acro{ODE}{operational distortion-energy}
\acro{ODR}{operational distortion-rate}
\acro{OFDM}{Orthogonal Frequency-Division Multiplexing}
\acro{omp}[OMP]{orthogonal matching pursuit}
\acro{OSMP}[OSMP]{orthogonal subspace matching pursuit}
\acro{OQAM}{offset quadrature amplitude modulation}
\acro{OQPSK}{offset QPSK}
\acro{OFDMA}{Orthogonal Frequency-division Multiple Access}
\acro{OPEX}{Operating Expenditures}
\acro{OQPSK/PM}{OQPSK with phase modulation}

\acro{PAM}{pulse amplitude modulation}
\acro{PAR}{peak-to-average ratio}
\acrodef{pdf}[PDF]{probability density function}                      
\acro{PDF}{probability density function}
\acrodef{p.d.f.}[PDF]{probability distribution function}
\acro{PDP}{power dispersion profile}
\acro{PMF}{probability mass function}                             
\acrodef{p.m.f.}[PMF]{probability mass function}
\acro{PN}{pseudo-noise}
\acro{PPM}{pulse position modulation}
\acro{PRake}{Partial Rake}
\acro{PSD}{power spectral density}
\acro{PSEP}{pairwise synchronization error probability}
\acro{PSK}{phase shift keying}
\acro{PD}{power density}
\acro{8-PSK}[$8$-PSK]{$8$-phase shift keying}
\acro{PPP}{Poisson point process}
\acro{PCP}{Poisson cluster process}
 
\acro{FSK}{Frequency Shift Keying}

\acro{QAM}{Quadrature Amplitude Modulation}
\acro{QPSK}{Quadrature Phase Shift Keying}
\acro{OQPSK/PM}{OQPSK with phase modulator }

\acro{RD}[RD]{raw data}
\acro{RDL}{"random data limit"}
\acro{ric}[RIC]{restricted isometry constant}
\acro{rict}[RICt]{restricted isometry constant threshold}
\acro{rip}[RIP]{restricted isometry property}
\acro{ROC}{receiver operating characteristic}
\acro{rq}[RQ]{Raleigh quotient}
\acro{RS}[RS]{Reed-Solomon}
\acro{RSC}[RSSC]{RS based source coding}
\acro{r.v.}{random variable}                               
\acro{R.V.}{random vector}
\acro{RMS}{root mean square}
\acro{RFR}{radiofrequency radiation}
\acro{RIS}{Reconfigurable Intelligent Surface}
\acro{RNA}{RiboNucleic Acid}
\acro{RRM}{Radio Resource Management}
\acro{RUE}{reference user equipments}
\acro{RAT}{radio access technology}
\acro{RB}{resource block}

\acro{SA}[SA-Music]{subspace-augmented MUSIC with OSMP}
\acro{SC}{small cell}
\acro{SCBSES}[SCBSES]{Source Compression Based Syndrome Encoding Scheme}
\acro{SCM}{sample covariance matrix}
\acro{SEP}{symbol error probability}
\acro{SG}[SG]{sparse-land Gaussian model}
\acro{SIMO}{single-input multiple-output}
\acro{SINR}{signal-to-interference plus noise ratio}
\acro{SIR}{signal-to-interference ratio}
\acro{SISO}{Single-Input Single-Output}
\acro{SMV}{single measurement vector}
\acro{SNR}[\textrm{SNR}]{signal-to-noise ratio} 
\acro{sp}[SP]{subspace pursuit}
\acro{SS}{spread spectrum}
\acro{SW}{sync word}
\acro{SAR}{specific absorption rate}
\acro{SSB}{synchronization signal block}
\acro{SR}{shrink and realign}

\acro{tUAV}{tethered Unmanned Aerial Vehicle}
\acro{TBS}{terrestrial base station}

\acro{uUAV}{untethered Unmanned Aerial Vehicle}
\acro{PDF}{probability density functions}

\acro{PL}{path-loss}

\acro{TH}{time-hopping}
\acro{ToA}{time-of-arrival}
\acro{TR}{transmitted-reference}
\acro{TW}{Tracy-Widom}
\acro{TWDT}{TW Distribution Tail}
\acro{TCM}{trellis coded modulation}
\acro{TDD}{Time-Division Duplexing}
\acro{TDMA}{Time Division Multiple Access}
\acro{Tx}{average transmit}

\acro{UAV}{Unmanned Aerial Vehicle}
\acro{uric}[URIC]{upper restricted isometry constant}
\acro{urict}[URICt]{upper restricted isometry constant threshold}
\acro{UWB}{ultrawide band}
\acro{UWBcap}[UWB]{Ultrawide band}   
\acro{URLLC}{Ultra Reliable Low Latency Communications}
         
\acro{wuric}[UWRIC]{upper weak restricted isometry constant}
\acro{wurict}[UWRICt]{UWRIC threshold}                
\acro{UE}{User Equipment}
\acro{UL}{UpLink}

\acro{WiM}[WiM]{weigh-in-motion}
\acro{WLAN}{wireless local area network}
\acro{wm}[WM]{Wishart matrix}                               
\acroplural{wm}[WM]{Wishart matrices}
\acro{WMAN}{wireless metropolitan area network}
\acro{WPAN}{wireless personal area network}
\acro{wric}[WRIC]{weak restricted isometry constant}
\acro{wrict}[WRICt]{weak restricted isometry constant thresholds}
\acro{wrip}[WRIP]{weak restricted isometry property}
\acro{WSN}{wireless sensor network}                        
\acro{WSS}{Wide-Sense Stationary}
\acro{WHO}{World Health Organization}
\acro{Wi-Fi}{Wireless Fidelity}

\acro{sss}[SpaSoSEnc]{sparse source syndrome encoding}

\acro{VLC}{Visible Light Communication}
\acro{VPN}{Virtual Private Network} 
\acro{RF}{Radio Frequency}
\acro{FSO}{Free Space Optics}
\acro{IoST}{Internet of Space Things}

\acro{GSM}{Global System for Mobile Communications}
\acro{2G}{Second-generation cellular network}
\acro{3G}{Third-generation cellular network}
\acro{4G}{Fourth-generation cellular network}
\acro{5G}{Fifth-generation cellular network}	
\acro{gNB}{next-generation Node-B Base Station}
\acro{NR}{New Radio}
\acro{UMTS}{Universal Mobile Telecommunications Service}
\acro{LTE}{Long Term Evolution}

\acro{QoS}{Quality of Service}
\end{acronym}
	
\newcommand{\SAR} {\mathrm{SAR}}
\newcommand{\WBSAR} {\mathrm{SAR}_{\mathsf{WB}}}
\newcommand{\gSAR} {\mathrm{SAR}_{10\si{\gram}}}
\newcommand{\Sab} {S_{\mathsf{ab}}}
\newcommand{\Eavg} {E_{\mathsf{avg}}}
\newcommand{\ft}{f_{\textsf{th}}}
\newcommand{\alphatf}{\alpha_{24}}

\title{
Ultra Reliable Low Latency Routing in LEO Satellite Constellations: A Stochastic Geometry Approach
}
\author{
Ruibo Wang, Mustafa A. Kishk, {\em Member, IEEE}, and Mohamed-Slim Alouini, {\em Fellow, IEEE}
\thanks{Ruibo Wang and Mohamed-Slim Alouini are with King Abdullah University of Science and Technology (KAUST), CEMSE division, Thuwal 23955-6900, Saudi Arabia. Mustafa A. Kishk is with the Department of Electronic Engineering, Maynooth University, Maynooth, W23 F2H6, Ireland. (e-mail: ruibo.wang@kaust.edu.sa; mustafa.kishk@mu.ie; slim.alouini@kaust.edu.sa). 
}
\vspace{-8mm}
}
\maketitle

\vspace{-0.8cm}

\begin{abstract}
In recent years, LEO satellite constellations have become envisioned as a core component of the next-generation wireless communication networks. The successive establishment of mega satellite constellations has triggered further demands for satellite communication advanced features: high reliability and low latency. In this article, we first establish a multi-objective optimization problem that simultaneously maximizes reliability and minimizes latency, then we solve it by two methods. According to the optimal solution, ideal upper bounds for reliability and latency performance of LEO satellite routing can be derived. Next, we design an algorithm for relay satellite subset selection, which can approach the ideal upper bounds in terms of performance. Furthermore, we derive analytical expressions for satellite availability, coverage probability, and latency under the stochastic geometry (SG) framework, and the accuracy is verified by Monte Carlo simulation. In the numerical results, we study the routing performance of three existing mega constellations and the impact of different constellation parameter configurations on performance. By comparing with existing routing strategies, we demonstrate the advantages of our proposed routing strategy and extend the scope of our research.
\end{abstract}


\section{Introduction}
In recent years, LEO satellite constellations have undergone explosive development and have gained an irreplaceable role in ultra-long distance communication operations, due to their extensive and seamless global coverage \cite{yue2023low,tian2021stochastic}. For long-distance routing, traditional small LEO satellite constellations maintain the store-and-forward communication mechanism: satellites temporarily store received data first and then forward it when they encounter the next hop within their line-of-sight (LoS) range \cite{lu2015complexity}. Due to the additional waiting time required during storage, communication often takes several minutes or even longer to complete. Recently, with the deployment of mega satellite constellations, the availability of satellites has been guaranteed like never before \cite{yuan2023joint}. Most likely, satellites can forward the message to another satellite as the next hop in a predetermined direction without waiting, thereby enabling real-time long-distance routing \cite{zhu2023timing,xv2023joint}.

\par
In addition to link availability, there is an increasing emphasis on other metrics for low-latency tolerant real-time LEO satellite communication services \cite{yue2022security}. Ultra-reliable low latency communications (URLLC), defined as a crucial metric by the 3rd Generation Partnership Project (3GPP), can function as an assessment criterion for terrestrial and near-ground networks. While satellite communications cannot meet the URLLC standards of ground networks due to significant path loss and long communication distances, high reliability and low latency can still serve as important objectives and guidelines in satellite routing design. Therefore, in this article, we design optimal routing strategies and evaluating the reliability, and latency performance of the proposed routing strategies.

\subsection{Related Works}
The design of satellite routing strategies and the strategies' performance analysis are widely discussed and highly regarded topics. However, existing researches often face a trade-off between the randomness of network topology and tractability. 

\par
Some studies have overlooked the unique dynamic topology of LEO satellite networks, as a trade-off for providing analytical results for latency and reliability. These literature design deterministic network topologies with pre-set starting and ending positions \cite{9348676_1,shen2020dynamic_2}, or referring to other existing deterministic network models \cite{geng2021agent_3,knight2011internet}, or even ignore the satellites' special spherical topology. However, since LEO satellites are not geosynchronous, the deterministic topology assumption may lead to inaccurate performance evaluation results and impractical routing strategies. Due to the high-speed movement of LEO satellites, a transmitting satellite that is communicating with the receiver may experience communication interruptions within a few minutes as it moves out of the receiver's line-of-sight (LoS) range \cite{Al-1}.

\par
Some studies consider the dynamic topology of LEO satellites and provide routing algorithms specifically designed for dynamic topologies, such as ant colony algorithm \cite{zhao2021multi_4} and particle swarm optimization algorithm \cite{8068282_5}. However, due to the heuristic nature of the algorithms, their stochastic performances, or even bounds of performances, are challenging to be expressed by analytical results \cite{hu2020directed,rabjerg2021exploiting}. 
The performance of these algorithms can only be assessed through numerical results, which means that any adjustment to a parameter's value may require a new simulation process, and in some cases, even algorithm redesign. If supported by analytical results, latency and reliability can be expressed as functions of parameters such as constellation altitude and the number of satellites. Researchers can intuitively observe the performance of routing algorithms as parameters vary.

\par
Therefore, we aim to find an analytical framework that possesses strong tractability while being able to capture characteristics of random topologies \cite{wang2022ultra}. SG is one of the most suitable mathematical tools for modeling and analyzing random topologies in large-scale networks \cite{al2021modeling}. In recent years, there have been numerous studies utilizing the SG analytical framework to investigate satellite performance and employ various models. Among these models, binomial point process (BPP) \cite{ok-2} is one of the most suitable one for modeling network devices in closed regions. Although there are some differences between the BPP and deterministic constellation models such as Walker's constellation, the performance estimations obtained from both models are very close \cite{ok-1,wang2022evaluating}.

\par
Some literature has investigated the reliability and latency of satellite communication under the SG framework. Reliability is measured by satellite availability \cite{wang2023reliability} and coverage probability \cite{ok-1,Al-1}. Coverage probability is the probability of successful signal demodulation at the receiver, which can be mathematically defined as the probability of the received signal-to-
interference plus noise ratio (SINR) greater than a threshold. As for latency, propagation latency and transmission latency are significantly affected by the random topology of the network \cite{rabjerg2021exploiting}, making them suitable for study within the framework of SG analysis. The propagation latency refers to the time delay required for an electromagnetic signal to propagate through space \cite{wang2022stochastic}. The transmission latency is the time delay required for the transmission of data packets at the achievable channel capacity and depends on the signal transmission rate and packet size.

\subsection{Contribution}

To our best knowledge, only our previous studies \cite{wang2023reliability} and \cite{wang2022stochastic} provide analytical expressions of availability and propagation latency in multi-hop satellite routing, respectively. However, our previous works do not consider the channel model, and thus, only focus on the randomness of network topology. Building upon this foundation, this article further explores the advantages of SG in analyzing stochastic channel fading. This allows for the estimation of system-level performance metrics such as routing coverage probability and transmission latency. The specific contributions are as follows.

\par
\begin{itemize}
\item Firstly, a greedy strategy is investigated by formulating a multi-objective (availability, coverage, and latency) optimization problem. Next, for an ideal scenario where satellite availability is guaranteed at any location, we employ two methods to solve the optimization problem and obtain corresponding optimal routing strategies. The communication performance of the idealized routing strategy can serve as  unattainable upper/lower bounds.
    
\item Based on insights into ideal scenario solutions, we propose an algorithm to select a subset of relay satellites in practical discrete scenarios. The complexity of the algorithm is analyzed.
    
\item Using the SG framework, we derive analytical expressions for the algorithm's availability, coverage probability, and latency. By introducing the automatic repeat request (ARQ) protocol, the routing transmission latency with ARQ protocol combining both reliability and latency is defined and its expression is derived as well. Considering the high computational complexity of two analytical expressions for latency, we provide two approximations.
    
\item Based on the numerical results of three realistic mega constellations (Starlink, Kuiper, and OneWeb), we validate the assumptions in this article and conduct a preliminary analysis of reliability and latency performance. Furthermore, we investigate the impact of different constellation parameter configurations on performance and compare the proposed routing strategy with other existing strategies.
\end{itemize}

\begin{figure}[htbp]
\begin{minipage}[t]{0.59\linewidth}
\centering
\includegraphics[width=0.98\linewidth]{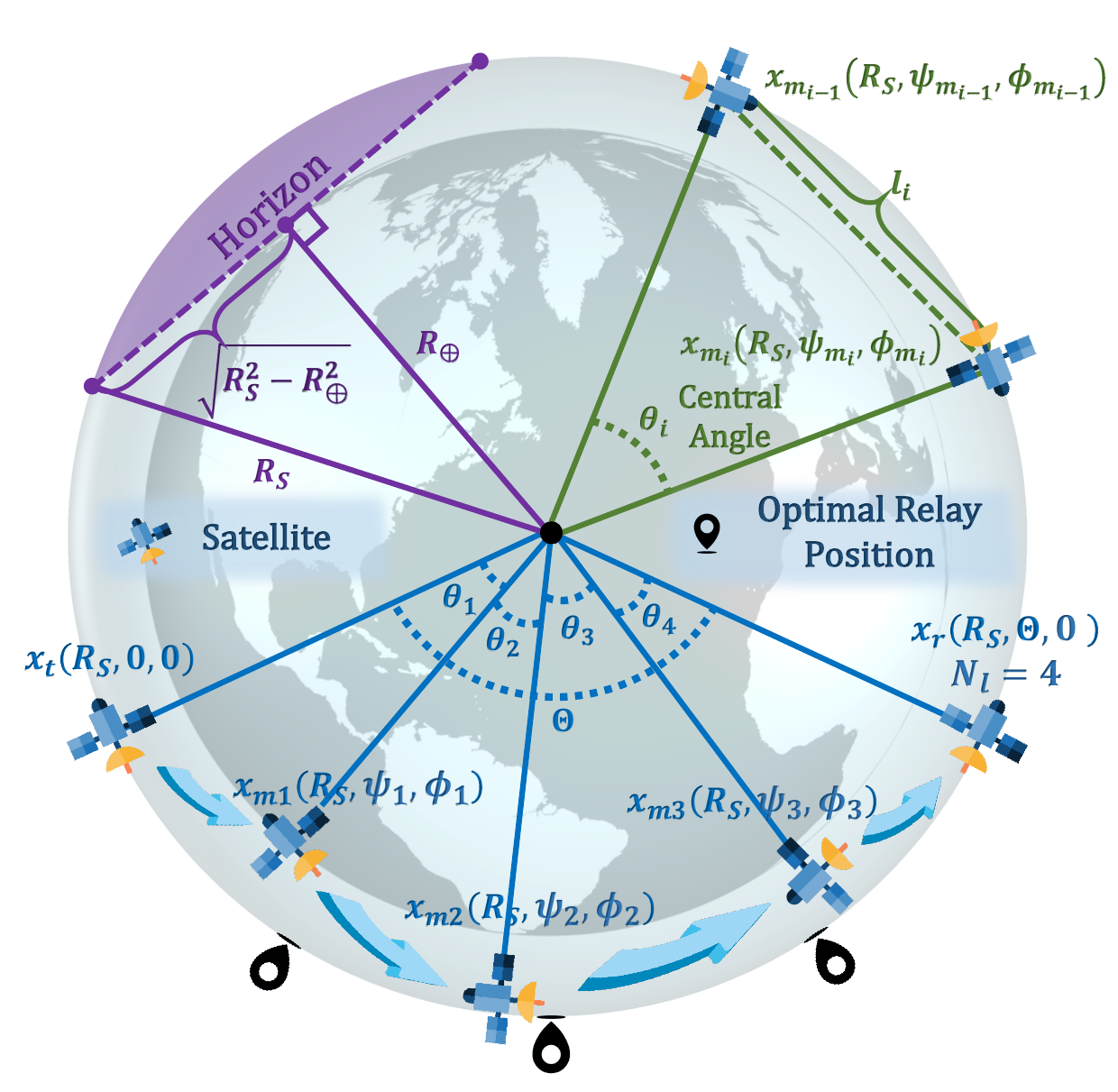}
\caption{Schematic diagram of the satellite network (the transmitting satellite $x_t$ sends a signal to the receiving satellite $x_r$ through satellite relays $x_{m_i}$).}
\label{fig:figure1}
\end{minipage}
\hfill
\begin{minipage}[t]{0.38\linewidth}
\centering
\includegraphics[width=0.98\linewidth]{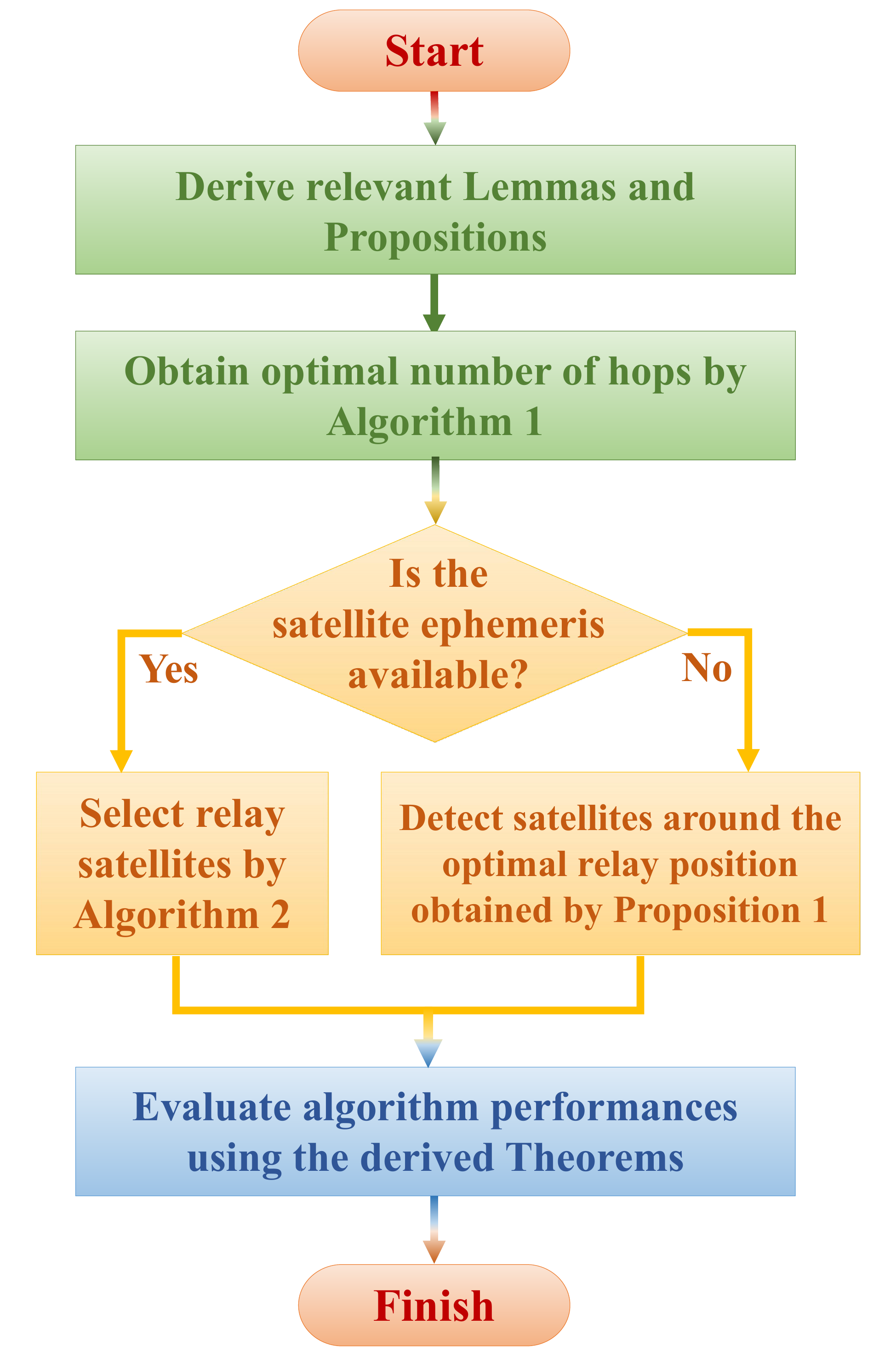}
\caption{Flowchart of the analytical  framework.}
\label{system}
\end{minipage}
\end{figure}

Finally, a flowchart of the analytical framework of this article is shown in Fig.~\ref{system}.


\section{System Model}
In this section, we first model satellite, route, and channel fading. Then the definitions of availability, coverage probability, and transmission latency are given. Based on these definitions, the original optimization problem is formulated. 

\subsection{Network Topology}\label{topology}
We consider a scenario where a satellite transmits the signal to another receiving satellite through multiple satellite relays, as shown in the bottom of Fig.~\ref{fig:figure1}. According to Slivnyak's theorem \cite{feller1991introduction}, the rotation of the coordinate system does not affect the satellite distribution\footnote{An explanation for this issue is that BPP exhibits independence and homogeneity. Specifically, for any location, the average number of satellites within a given region is directly proportional to the area of this region \cite{wang2023terrain}. As the distribution of BPP is independent of position, the distribution of satellites should appear uniform from any point of view \cite{lou2023coverage}.}. Without loss of generality, we take the Earth center as the origin, the coordinate of the transmitter in the spherical coordinate system as $x_t (R_s,0,0)$, and the coordinate of the receiver as $x_r (R_s,\Theta,0)$. $R_s$ is the altitude of the LEO satellites plus the radius of the earth, which represents the radius of the sphere where satellites are located. $\Theta$ represents the polar angle of the receiver, and the azimuth angles of the transmitter and receiver are both $0$.

\par
We consider $N_s$ relay satellites to be distributed on a sphere around Earth, whose locations form a spherical homogeneous BPP denoted as $\mathcal{X}=\{x_1,x_2,...,x_{N_s}\}$, where $x_i(R_s,\psi_i,\phi_i)$ is the location of the $i^{th}$ satellite, $\psi_i$ and $\phi_i$ are the polar and azimuth angles of the $i^{th}$ satellite, respectively. We define a transmission from one communication device to another as a hop. A route with $N_l$ hops is expressed as $\mathcal{M}_{|N_l}= \{m_1,...,m_{N_l-1}\}$, such that $\{x_{m_1},\dots,x_{m_{N_l-1}}\} \in \mathcal{X}$, where $m_{i-1}$ and $m_i$ are the indices of the two satellites corresponding to the $i^{th}$ hop, $2 \leq i \leq N_l-1$. The hop from the transmitter at $x_t (R_s,0,0)$ to the first relay satellite at $x_{m_1} (R_s,\psi_{m_1},\phi_{m_1})$ is the first hop, and the hop from the last relay satellite at $x_{m_{N_l-1}}(R_s,\psi_{m_{N_l-1}},\phi_{m_{N_l-1}})$ to the receiver at $x_r (R_s,\Theta,0)$ is the last hop. For the sake of representation, $m_0$ and $m_{N_l}$ are labeled as indices of the transmitter $x_t$ and receiver $x_r$. Then, we denote $l_i = d \left( \psi_{m_{i-1}},\phi_{m_{i-1}} ; \psi_{m_i}, \phi_{m_i} \right)$ as the Euclidean distance of the $i^{th}$ hop, where $d \left( \psi_1,\phi_1 ; \psi_2, \phi_2 \right)$ represents the Euclidean distance between  $(R_s,\psi_1,\phi_1)$ and  $(R_s,\psi_2,\phi_2)$:
\begin{equation}\label{d_12}
\begin{split}
    d \left( \psi_1,\phi_1 ; \psi_2, \phi_2 \right) = \sqrt{2 R_s^2 \big(1 - \cos{\psi_1}\cos{\psi_2} - \sin{\psi_1}\sin{\psi_2}\cos\left(\phi_1-\phi_2 \right) \big)}.
\end{split}
\end{equation}
In addition, we denote $\theta_i=2\arcsin\frac{l_i}{2R_s}$ as the central angle of the $i^{th}$ hop, as depicted in the upper right part of Fig.~\ref{fig:figure1}. The definition of the central angle is given as follows.

\begin{definition}[Central Angle]
    For two given points $x_A$ and $x_B$, the central angle between $x_A$ and $x_B$ is the angle formed by the ray from the Earth's center to $x_A$, and the ray from the Earth's center to $x_B$.
\end{definition}

\subsection{Channel Fading Model}
Based on existing research, we consider the channel fading model of satellite communication to follow the free space fading model with ideal feeder links \cite{na2021performance}. The received power $\rho_i^r$ of the $i^{th}$ hop under the channel fading model can be expressed as follows \cite{kaur2019analysis}, 
\begin{equation} \label{rho_i}
    \rho_i^r = \rho^t \, G \left( \frac{\lambda}{4\pi l_i} \right)^2 W,
\end{equation}
where $\rho^t$, $G$, $\lambda$, and $W$ denote the transmission power, antenna gain, wavelength, and small scale fading, respectively.

\par
Unlike satellite-terrestrial links, the impact of the multipath effect in the ground clutter layer and rain attenuation in the atmosphere  on inter-satellite links (ISLs) can be neglected \cite{bitragunta2020best}. Due to the high mobility of satellites and the application of extremely narrow beams, the small scale fading of the ISL is mainly caused by the pointing error. Therefore, we consider the envelope of the small scale fading power following the pointing error model in \cite{ata2022performance}. Given that the deviation angle of the beam is equal to $\theta_d$, the conditional probability density function (PDF) of the small scale fading gain $W$ can be written as \footnote{The deviation in the vertical direction and the deviation in the horizontal direction are assumed to have the same jitter variance, corresponding to the case where the pointing error is maximum.}
\begin{equation}
\label{pointing_error}
    f_{W|\theta_d}\left ( w \right ) = \frac{\eta_s^2 w^{\eta_s^2-1 } \cos\theta_d}{A_0^{\eta_s^2}}, \ 0 \leq w \leq A_0,
\end{equation}
where $\eta_s$ and $A_0$ are parameters of the pointing error. We consider the deviation angle of inter-satellite links to follow Rayleigh distribution with variance $\varsigma^2$ \cite{ata2022performance}, 
\begin{equation}\label{fthetad}
    f_{\theta_d}\left ( \theta_d \right )=\frac{\theta_d}{\varsigma^2}\exp\left ( -\frac{\theta_d^2}{2\varsigma^2} \right ), \ \theta_d\geq 0.
\end{equation}
So far, the distribution of inter-satellite links' deviation angle $\theta_d$ is still to be studied. Fortunately, we will show that the theorems provided in this article will apply to any distribution of $\theta_d$.

\subsection{Metrics Definition}
In this subsection, we introduce the definitions of three metrics: availability, coverage probability, and latency. The first two metrics are used to measure reliability. 
\par

\begin{definition}[Availability]\label{def2}
\hspace*{\fill}
\begin{itemize}
    \item If the link between two communication devices of a hop is not blocked by the Earth, this hop is available \cite{ok-2}. 
    \item A route is available if every hop in this route is available. 
\end{itemize}
\end{definition}
As indicated in the upper left part of Fig.~\ref{fig:figure1}, a sufficient and necessary condition for single-hop availability is satisfying the following restriction on single-hop distance: 
\begin{equation}\label{def_avail}
    l_i \leq 2\sqrt{R_s^2 - R_\oplus^2} \, ,
\end{equation}
where $R_\oplus$ denotes the radius of the Earth.

\begin{definition}[Coverage Probability]\label{def_cover}
\hspace*{\fill}
\begin{itemize}
    \item The conditional coverage probability is the probability that the signal-to-noise ratio (SNR) at a single-hop's receiver reaches a particular threshold under the condition that the distance of this hop is known \cite{lou2023cover}. 
    \item The routing coverage probability is the probability that the SNR at every hop's receiver reaches the threshold. 
\end{itemize}
\end{definition}
Given that the distance of the $i^{th}$ hop is $l_i$, the conditional coverage probability of the $i^{th}$ hop is defined as 
\begin{equation}\label{condicover}
    P^C_{\rm cond} (l_i) = \mathbb{P} \left[ {\rm SNR}_i= \frac{\rho_{i}^r}{\sigma^2} \geq \gamma | \, l_i \right],
\end{equation}
where $\gamma$ is the coverage threshold, and $\sigma^2$ is the noise power. We consider that the communication is noise-dominated, and the power of co-channel interference is negligible. This is reasonable because of the satellite's narrow beam transmission mode and sufficient bandwidth to achieve orthogonal frequency allocation \cite{okati2020stochastic}. This assumption will be further verified in numerical results.

\begin{definition}[Latency] The routing transmission latency is the sum of the average packet transmission latency at the achievable data rate of each hop.
\end{definition}
From the definition, the latency can be written as
\begin{equation}\label{deflatency}
    T_{\rm{tx}} = \sum_{i=1}^{N_l} \frac{\varpi }{B \log_{2}\left(1+{\rm{SNR}}_i \right)},
\end{equation}
where $\varpi$ is the packet size, $B$ denotes the channel bandwidth, and ${\rm{SNR}}_i$ is defined in (\ref{condicover}). The propagation latency is ignored in the optimization, which will be explained in Sec.~\ref{sec5-1}.

\subsection{Problem Formalization}
Therefore, the original optimization problem $\mathscr{P}_1$ includes three optimization objectives: 
\begin{subequations} 
	\begin{alignat}{2}
		\mathscr{P}_1: &\underset{N_l,\mathcal{M}_{|N_l}}{\mathrm{maximize}} \ \ \ \ \  & \prod_{i=1}^{N_l} \mathbbm{1} \Big\{& {l_i}  \leq 2\sqrt{R_s^2 - R_\oplus^2} \Big\}, \label{opt1-1}\\
		&\underset{N_l,\mathcal{M}_{|N_l}}{\mathrm{maximize}} \ & & \prod_{i=1}^{N_l}  P^C_{\rm cond}(l_i), \label{opt1-2} \\
		&\underset{N_l,\mathcal{M}_{|N_l}}{\mathrm{minimize}} \ &   \sum_{i=1}^{N_l} & \frac{\varpi }{B \log_{2} (1+{\rm{SNR}}_i) }, \label{opt1-3} 
	\end{alignat}
\end{subequations}
where optimization objectives in (\ref{opt1-1}), (\ref{opt1-2}), and (\ref{opt1-3}) represent link availability, link coverage probability, and latency, respectively. $\mathbbm{1}\{ \cdot \}$ denotes an indicator function, $\mathbbm{1}\{ \cdot \}=1$ when condition $\cdot$ is satisfied, $\mathbbm{1}\{ \cdot \}=0$ otherwise. Recall that $N_l$ is the number of hops in a link, and $\mathcal{M}_{|N_l}= \{m_1,...,m_{N_l-1}\}$ records the indices of relay satellites.

\par
However, solving the original optimization problem $\mathscr{P}_1$ faces three challenges: $\mathscr{P}_1$ is a (i) multi-objective and (ii) discrete optimization problem, (iii) $l_i$ and small scale fading are random variables. Note that the distribution of satellites follows BPP, so the topology of the network is random. Even with the same routing strategy, the single-hop distance $l_i$ in each BPP realization is different and random. To address the challenge (i), we adopt two different methods in Sec.~\ref{section3} and Sec.~\ref{section4}.

\section{Optimization with Constraints (Method I)}\label{section3}
In this section, we first solve the optimization problem in an ideal scenario where satellites are available at any location. Then, we propose an algorithm to transit the idea scenario to a discrete scenario where satellite distribution follows the BPP. Finally, analytical expressions for three metrics under the implementation of the proposed algorithm are provided.

\subsection{Ideal Scenario Solution}\label{sec3-1}
In this subsection, we first convert availability and coverage probability from optimization objectives to constraints and then make $\mathscr{P}_1$ a single objective optimization problem with constraints. However, the optimization of hop number and relay satellite selection are still discrete problems. In addition, $l_i$ and $W$ are random variables, and it is more challenging to solve optimization problems with random variables. 

\par
Therefore, to compute performance bounds, we assume that satellites are available at any given location. Under this ideal assumption, $l_i$ is no longer a random variable, and finding locations of relay satellites is also transformed from discrete subset selection to a continuous optimization problem. Consider that $N_l$ and $\mathcal{M}_{|N_l}$ are independent of small scale fading $W$, we take the mean value of $W$ fading and $\mathscr{P}_1$ can be rewritten as follows:
\begin{subequations} 
	\begin{alignat}{2}
		\mathscr{P}_2: \underset{N_l}{\mathrm{minimize}} \ \underset{\mathcal{M}_{|N_l}}{\mathrm{minimize}} \ \ \ &  \sum_{i=1}^{N_l} \frac{\varpi }{B \log_{2}\left(1+\mathbb{E}_W [ {\rm{SNR}}_i] \right)}, \label{opt2-1}\\
		\mathrm{subject \ to}  \ \ \ \ \ \   & \prod_{i=1}^{N_l} P^C_{\rm cond}(l_i) \geq 1 - \varepsilon, \label{opt2-2} \\
		& \ l_i \leq 2\sqrt{R_s^2 - R_\oplus^2}, \ \forall i, \label{opt2-3}
	\end{alignat}
\end{subequations}
where $\varepsilon$ in (\ref{opt2-2}) is the tolerable probability of communication interruption/not being covered.

\par
$\mathscr{P}_2$ is still a discrete optimization problem because $N_l$ is an integer and it is unreasonable for $N_l$ to be a non-integer value. Fortunately, it is not computationally expensive to find the optimal number of hops even by exhaustive search. Therefore, we solve $\mathscr{P}_2$ by two steps: continuous optimization of $\mathcal{M}_{|N_l}$ ($\mathscr{P}_{2-1}$) and discrete solution of $N_l$ ($\mathscr{P}_{2-2}$). $\mathscr{P}_{2-1}$ aims to find the optimal relay positions in the ideal scenario with a fixed number of hops:
\begin{subequations} 
	\begin{alignat}{2}
		\mathscr{P}_{2-1}: \underset{\mathcal{M}_{|N_l}}{\mathrm{minimize}} \ \ \ &  \sum_{i=1}^{N_l} \frac{\varpi }{B \log_{2}\left(1+\mathbb{E}_W[{\rm{SNR}}_i] \right)}, \label{opt3-1}\\
		\mathrm{subject \ to} \ \ & l_i \leq 2\sqrt{R_s^2 - R_\oplus^2}. \label{opt3-2}
	\end{alignat}
\end{subequations}
The following proposition can provide a potential solution to $\mathscr{P}_{2-1}$. 

\begin{proposition}\label{prop1}
For a given number of hops $N_l$, there exists a solution to $\mathscr{P}_{2-1}$ in the feasible region, when $2 N_l \arccos\frac{R_\oplus}{R_s} \geq \Theta$ is satisfied. The positions of the relay satellite of this solution are $\left\{ \left(R_s,\frac{\Theta}{N_l},0\right), \left(R_s,\frac{2\Theta}{N_l},0\right), \dots, \left(R_s,\frac{(N_l-1)\Theta}{N_l},0\right) \right\}$, and the corresponding local minimum latency can be written as $N_l \ T_{{\rm tx},1}^* \left(2R_s \sin \frac{\Theta}{2 N_l} \right)$, where $T_{{\rm tx},1}^* \left(l_i\right)$ is defined as
\begin{equation}\label{T_tx}
\begin{split}
    T_{{\rm tx},1}^* \left(l_i\right) = \frac{\varpi}{B \log_2 \left(1+\rho^t \, G \left( \frac{\lambda}{4\pi l_i} \right)^2 \frac{A_0 \eta_s^2}{1 + \eta_s^2} \left( 1 - \varsigma^2 \right) \sigma^{-2} \right) }.
\end{split}
\end{equation}
\begin{proof}
See Appendix~\ref{app:prop1}.
\end{proof}
\end{proposition}

Although the solution given in Proposition~\ref{prop1} is not necessarily optimal and unique, it provides important insights into the relay satellite selection in discrete scenarios. So far, we get an idea of the potential optimal satellite relay positions under a fixed number of hops. Next, $\mathscr{P}_{2-2}$ aims to find the optimal number of hops on the premise of maintaining the optimal relay distribution proposed in Proposition~\ref{prop1}:
\begin{subequations} 
	\begin{alignat}{2}
		\mathscr{P}_{2-2}: \ \ &\underset{N_l}{\mathrm{minimize}} \ \ \  N_l \ T_{{\rm tx},1}^* \left(2R_s \sin \frac{\Theta}{2 N_l} \right), \label{opt4-1}\\
		& \mathrm{subject \ to} \ \  P^C_{\rm cond}\left(2R_s \sin \frac{\Theta}{2N_l} \right) \geq \sqrt[N_l]{1 - \varepsilon}, \label{opt4-2} \\
		& \ \ \ \ \ \ \ \ \ \ \ \ \ \ \ 2 N_l \arccos\frac{R_\oplus}{R_s} \geq \Theta, \label{opt4-3}
	\end{alignat}
\end{subequations}
where $T_{\rm tx}^* \left(N_l\right)$ is defined in (\ref{T_tx}), and the conditional coverage probability is given in the following lemma. The inequality (\ref{opt4-2}) ensures the probability that all of the $N_l$ hops reach the coverage threshold is greater than $1 - \varepsilon$, where $\varepsilon$ has the same meaning as in (\ref{opt2-2}), and $2R_s \sin \frac{\Theta}{2N_l}$ is the Euclidean distance of a single-hop.

\begin{lemma}\label{lemma1}
    Given that the distance of the $i^{th}$ hop is $l_i$, the conditional coverage probability is written as
\begin{equation}\label{PCcond}
   P^C_{\rm cond} (l_i) = 1 - \varsigma^2 - \left( 1 - \varsigma^2 \right)\left(\frac{\gamma \sigma^2 \left( 4\pi l_i \right)^2}{A_0 \rho^t \, G \lambda^2}\right)^{\eta_s^2}.
\end{equation}
\begin{proof}
    See Appendix~\ref{app:lemma1}.
\end{proof}
\end{lemma}

Lemma~\ref{lemma1} shows that $P^C_{\rm cond} (l_i) < 1 - \varsigma^2$ for $\forall \, l_i$. Combined with (\ref{opt4-2}), the inequality $\left( 1 - \varsigma^2 \right)^{N_l} > 1 - \varepsilon$ needs to be satisfied, from which we can provide an upper bound for $N_l$. Based on the above analysis, the following proposition summarizes the cases of the solution of $\mathscr{P}_{2-2}$.

\begin{proposition}\label{prop2}
    The solution of optimization problem $\mathscr{P}_{2-2}$ have the following disjoint cases:
    \begin{itemize}
        \item Case 1: If $\frac{\Theta}{2\arccos(R_{\oplus}/R_s)} \geq \frac{\ln (1-\varepsilon)}{\ln \left( 1 - \varsigma^2 \right)}$, $\mathscr{P}_{2-2}$ has no feasible solutions.
        \item Case 2: If no integer $N_l$ satisfies both $\frac{\Theta}{2\arccos(R_{\oplus}/R_s)} \leq N_l < \frac{\ln (1-\varepsilon)}{\ln \left( 1 - \varsigma^2 \right)}$ and constraint condition (\ref{opt4-2}), $\mathscr{P}_{2-2}$ has no feasible solutions.
        \item Case 3: Traverse every $N_l$ that satisfies both $\frac{\Theta}{2\arccos(R_{\oplus}/R_s)} \leq N_l < \frac{\ln (1-\varepsilon)}{\ln \left( 1 - \varsigma^2 \right)}$ and constraint condition (\ref{opt4-2}), and the $N_l$ that minimizes $T_{\rm tx}^* \left(N_l\right)$ is the solution to $\mathscr{P}_{2-2}$.
    \end{itemize}
\end{proposition}


\subsection{Discrete Scenario Algorithms}
In order to map the optimal relay positions in the ideal scenario to the selected satellite subset in the discrete scenario, an intuitive idea is to find satellites in $\mathcal{X}$ that are closest to the ideal optimal relay positions. However, this mapping method causes the average distance of each hop in BPP to be larger than that in the ideal scenario. A simple explanation is that the selected relay satellite in the discrete point processes will more or less deviate from the shortest inferior arc. Therefore, we introduce a term called distance scaling factor to measure the ratio of the average distance in the discrete scenario to the distance in the ideal scenario. 
\begin{lemma}\label{lemma2}
    The distance scaling factor of the first or last hop is
    \begin{equation}
    \begin{split}
        \alpha^{(1)} (\theta_i) &= \frac{N_s}{8\pi R_s \sin\frac{\theta_i}{2}} \int_0^{2\pi} \int_0^{\pi} \sin\xi \left( \frac{ 1 + \cos\xi }{2} \right)^{N_s-1} d\left(\theta_i,0;\xi,\varphi \right) \mathrm{d}\xi \mathrm{d}\varphi, \ \ i=\{1,N_l\},
    \end{split}
    \end{equation}
    where $\theta_i$ is the central angle of the $i^{th}$ hop, and $d\left(\theta_i,0;\xi,\varphi \right)$ is given in (\ref{d_12}).
    
    \par
    The distance scaling factor of a middle hop $\alpha^{(2)} \left(\theta_i\right), \, 2\leq i \leq N_l-1$ can be approximately estimated by $\alpha^{(2)} \left(\theta_i\right) = 2\alpha^{(1)} \left(\theta_i\right) - 1$ (called additive evaluation) or by $\alpha^{(2)} \left(\theta_i\right) = \left(\alpha^{(1)} \left(\theta_i\right)\right)^2$ (called multiplicative evaluation).
\begin{proof}
    See Appendix~\ref{app:lemma2}.
\end{proof}
\end{lemma}

\par
In Lemma~\ref{lemma2}, we propose two estimation for $\alpha^{(2)} \left(\theta_i\right)$. We can use either estimation for the rest of this article. In fact, for a massive LEO satellite constellation, $\alpha^{(1)} \left(\theta_i\right)$ and $\alpha^{(2)} \left(\theta_i\right)$ are close to 1. In this case, the difference between multiplicative estimation and additive estimation is not significant, since $\left(\alpha^{(1)} \left(\theta_i\right)\right)^2 - \left( 2\alpha^{(1)} \left(\theta_i\right) -1 \right) = \left(\alpha^{(1)} \left(\theta_i\right) - 1\right)^2 \rightarrow 0$. Based on distance scaling factors $\alpha^{(1)} \left(\theta_i\right)$ and $\alpha^{(2)} \left(\theta_i\right)$, we propose a relay satellite selection algorithm to solve $\mathscr{P}_2$ in the discrete scenario. Similar to the solving procedure in the ideal scenario, the algorithm is divided into two steps: finding the optimal number of hops and searching for the best relay satellites. 

\begin{algorithm}[!ht] 
    \caption{Exhaustive Search for Optimal Number of Hops}
	\label{alg1} 
	\begin{algorithmic} [1]
		\STATE \textbf{Input}: Tolerable probability of communication interruption $\varepsilon$.

        \STATE \textbf{Initiate} $N_l^* \leftarrow 0$ and $T_{\min} \leftarrow \infty$.
		\IF{$\frac{\Theta}{2\arccos(R_{\oplus}/R_s)} \geq \frac{\ln (1-\varepsilon)}{\ln \left( 1 - \varsigma^2 \right)}$}
        \STATE \textbf{Exit} the algorithm and \textbf{output} $N_l^*$.
        \ENDIF

        \STATE $N_l \leftarrow \left\lceil \frac{\Theta}{2\arccos(R_{\oplus}/R_s)} \right\rceil$.
        
		\WHILE{$\frac{\Theta}{2\arccos(R_{\oplus}/R_s)} \leq N_l < \frac{\ln (1-\varepsilon)}{\ln \left( 1 - \varsigma^2 \right)}$}
        \STATE $N_l \leftarrow N_l + 1$, $l_{\rm Alg1} \leftarrow 2R_s \sin\frac{\Theta}{2N_l}$.
        
		\IF{$\left( P^C_{\rm cond}\left( \alpha^{(1)}\left( \frac{\Theta}{N_l} \right) l_{\rm Alg1} \right) \right)^2 \times \left( P^C_{\rm cond}\left( \alpha^{(2)}\left( \frac{\Theta}{N_l} \right) l_{\rm Alg1} \right) \right)^{N_l-2} \geq 1 - \varepsilon$ and \\ $\alpha^{(2)}\left( \frac{\Theta}{N_l} \right) l_{\rm Alg1} < 2 \sqrt{R_s^2 - R_{\oplus}^2}$ }
  
        \STATE $T_{\rm Alg1} \leftarrow 2 T_{{\rm tx},1}^* \left(\alpha^{(1)}\left( \frac{\Theta}{N_l} \right) l_{\rm Alg1}\right) + \left(N_l - 2\right) T_{{\rm tx},1}^* \left(\alpha^{(2)}\left( \frac{\Theta}{N_l} \right) l_{\rm Alg1}\right)$.
        
        \IF{$T_{\min} \geq T_{\rm Alg1}$}
        
        \STATE $T_{\min} \leftarrow T_{\rm Alg1}$, $N_l^* \leftarrow N_l$.
        \ENDIF
        \ENDIF
        
		\ENDWHILE
		
		\STATE \textbf{Output}: Optimal number of hops $N_l^*$.
	\end{algorithmic}
\end{algorithm}	

In Algorithm~\ref{alg1}, $P^C_{\rm cond}(l_i)$ and $T_{{\rm tx},1}^*(l_i)$ are defined in (\ref{PCcond}) and (\ref{T_tx}), respectively. Step (3)-(5), step (7)-(9), and step (10)-(13) correspond to the verification in case 1, case 2, and exhaustive search in case 3 from Proposition~\ref{prop2}, respectively. When $\mathscr{P}_2$ has no feasible solutions, $N_l^*$ is equal to $0$. $\lceil \cdot \rceil$ in step (6) represents rounding up to an integer. Step (9) in algorithm~\ref{alg1} verifies whether the expected routing scheme is reliable by coverage probability and availability. $T_{\rm Alg1}$ in step (10) records the average latency given that the number of hops is $N_l$. Algorithm~\ref{alg1} has a low computational complexity because it only involves a single loop in step (7).

\begin{algorithm}[!ht]  
	\caption{Relay Satellite Selection Algorithm}
	\label{alg2}
	\begin{algorithmic} [1]
		
		\STATE \textbf{Input}: Locations of satellites $\mathcal{X}$.
        
        \STATE Obtain optimal number of hops $N_l^*$ through Algorithm~\ref{alg1}: exhaustive search for optimal number of hops.
        
		\FOR{$i = 1 : N_l^*-1$}
		\STATE $m_i \leftarrow \arg {\min\limits_{1\leq n \leq N_s}} \; d\left( i \times \frac{ \Theta}{N_l},0;\psi_n,\phi_n \right)$.
        \STATE $\mathcal{B}_i \leftarrow \mathbbm{1} \left\{ d\left( \psi_{m_{i-1}},\phi_{m_{i-1}}; \psi_{m_i},\phi_{m_i} \right) > 2\sqrt{R_s^2 - R_{\oplus}^2} \right\}$.
		\ENDFOR
		\STATE $\mathcal{B}_{N_l^*} \leftarrow \mathbbm{1} \left\{ d\left( \psi_{m_{N_l^*-1}},\phi_{m_{N_l^*-1}}; \Theta,0 \right) > 2\sqrt{R_s^2 - R_{\oplus}^2} \right\}$.
		\STATE $\mathcal{M}_{|N_l^*} \leftarrow \{ m_1,m_2,...,m_{N_l^*-2},m_{N_l^*-1} \}. $
		
		\STATE \textbf{Output}: IDs of relay satellites in the route $\mathcal{M}_{|N_l^*}$ and Boolean variables $\mathcal{B}_i,1\leq i \leq N_l^*$ that imply unavailability.
	\end{algorithmic}
\end{algorithm}

In Algorithm~\ref{alg2}, step (4) searches the relay satellites in $\mathcal{X}$ that are closest to optimal relay positions in the ideal scenario. Steps (5) and (7) verify whether each hop of the route is interrupted because of Earth blockage. $\mathbbm{1} \{ \cdot \}$ is an indicator function. When the condition in $\mathbbm{1} \{ \cdot \}$ satisfies, $\mathbbm{1} \{ \cdot \} = 1$, otherwise $\mathbbm{1} \{ \cdot \}=0$. $d\left( \psi_{m_{i-1}},\phi_{m_{i-1}}; \psi_{m_i},\phi_{m_i} \right)$ is defined in (\ref{d_12}). When $\mathcal{B}_i = 1$, we need to find one or more extra relay satellites between the satellites with IDs $m_{i-1}$ and $m_i$ \footnote{In numerical results, we choose the minimum deflection angle strategy proposed in \cite{wang2022stochastic} to find extra relay satellites. In the numerical simulation, the probability of unavailability of the relay satellite is less than $0.1\%$ through a reasonable constellation design. Subsequent numerical results will demonstrate that when rare cases of satellite unavailability is considered, the analytical expressions provided still perfectly match the average performance obtained from the Monte Carlo simulation.}. The complexity of Algorithm~\ref{alg2} is significantly higher than Algorithm~\ref{alg1}. The computational complexity of Algorithm~\ref{alg2} mainly concentrates on step (4). If the calculation of Euclidean distance $d\left( \psi_{m_{i-1}},\phi_{m_{i-1}}; \psi_{m_i},\phi_{m_i} \right)$ is taken as a unit of computational complexity, the computational complexity of Algorithm~\ref{alg2} is $\mathcal{O}\left(N_s N_l^*\right)$.

\subsection{Metrics Analysis}
This subsection analyzes the performance of Algorithm~\ref{alg2} through reliability and latency-related metrics. In the beginning, we introduce an intermediate variable called single-hop central angle distribution, which has a tight relationship to the distance scaling factor.

\begin{definition}[Central Angle Distribution]
For two given reference positions $x_A$ and $x_B$ that are not part of point process $\mathcal{\widetilde{X}}$, their nearest neighbors in $\mathcal{\widetilde{X}}$ are denoted as $x_C$ and $x_D$, respectively. 
\begin{itemize}
    \item The distribution of the central angle between $x_A$ and $x_D$ (or $x_B$ and $x_C$) is called type-I central angle distribution. The central angle distribution of the first or last hop follows the type-I central angle distribution.
    \item The distribution of the central angle between $x_C$ and $x_D$ is called type-II central angle distribution. The central angle distribution of a middle hop follows the type-II central angle distribution. 
\end{itemize}
\end{definition}

According to Definition~\ref{def2}, the probability that the route is available is equal to the product of the probability that the distance of each hop is less than $2\sqrt{R_s^2-R_{\oplus}^2}$. Therefore, availability can be expressed by the cumulative distribution function (CDF) of central angle distribution.

\begin{theorem}\label{theorem1}
The probability that a route $\mathcal{M}_{|N_l^*}$ obtained by Algorithm~\ref{alg2} is available is
\begin{equation}
\begin{split}
    P^A & = \left( F_{\theta_c^{(1)}} \left(2\arcsin\frac{\sqrt{R_s^2-R_{\oplus}^2}}{R_s}\right) \right)^2 \times \left(  F_{\theta_c^{(2)}} \left(2\arcsin\frac{\sqrt{R_s^2-R_{\oplus}^2}}{R_s}\right) \right)^{N_l-2},
\end{split}
\end{equation}
where $F_{\theta_c^{(1)}} \left(\Xi\right)$ is the CDF of type-I central angle distribution,
\begin{equation}
\begin{split}
    & F_{\theta_c^{(1)}} \left(\Xi\right) = \int_0^{2\pi} \int_0^{\Xi} \frac{N_s \sin\theta}{2^{N_s+1} \pi} \left( 1 + \cos\left( 2\arcsin \frac{d(\theta,\phi;\frac{\Theta}{N_l},0)} {2R_s} \right) \right)^{N_s-1} \mathrm{d}\theta \mathrm{d}\phi,
\end{split}
\end{equation}
and $d(\theta,\phi;\frac{\Theta}{N_l},0)$ is defined in (\ref{d_12}). $F_{\theta_c^{(2)}} \left(\Xi\right)$ is the CDF of type-II central angle distribution, which can be derived from the CDF of type-I central angle distribution,
\begin{equation}
\begin{split}
    F_{\theta_c^{(2)}} \left(\Xi\right) = F_{\theta_c^{(1)}} \left(2 \arcsin \frac{\sin(\Xi/2)}{\alpha^{(1)}\left( \Theta/N_l \right)}\right).
\end{split}
\end{equation}
\begin{proof}
    See Appendix~\ref{app:theorem1}.
\end{proof}
\end{theorem}

In addition to availability, the routing coverage probability in Definition~\ref{def_cover} is also closely related to the PDF of central angle distribution.

\begin{theorem}\label{theorem2}
    The routing coverage probability of route $\mathcal{M}_{|N_l^*}$ obtained by Algorithm~\ref{alg2} is
\begin{equation}\label{PCrout}
\begin{split}
    P^C_{\rm rout} &= \left(\int_0^{\pi} f_{\theta_c^{(1)}}(\theta) P^C_{\rm cond} \left(2R_s\sin\frac{\theta}{2}\right) \mathrm{d}\theta\right)^2  \times \left(\int_0^{\pi} f_{\theta_c^{(2)}}(\theta) P^C_{\rm cond} \left(2R_s\sin\frac{\theta}{2}\right) \mathrm{d}\theta\right)^{N_l-2},
\end{split}
\end{equation}
where conditional coverage probability $P^C_{\rm cond} \left(2R_s\sin\frac{\theta}{2}\right)$ is defined in (\ref{condicover}), $f_{\theta_c^{(1)}}(\theta)$ is the PDF of type-I central angle distribution,
\begin{equation}\label{fthetaPDF1}
\begin{split}
    & f_{\theta_c^{(1)}} \left(\theta\right)  = \int_0^{2\pi} \frac{N_s \sin\theta}{2^{N_s+1} \pi}  \left( 1 + \cos\left( 2\arcsin \frac{d(\theta,\phi;\frac{\Theta}{N_l},0)} {2R_s} \right) \right)^{N_s-1} \mathrm{d}\phi,
\end{split}
\end{equation}
$d(\theta,\phi;\frac{\Theta}{N_l},0)$ is defined in (\ref{d_12}), and $f_{\theta_c^{(2)}}(\theta)$ is the PDF of type-II central angle distribution,
\begin{equation}\label{fthetaPDF2}
\begin{split}
    f_{\theta_c^{(2)}} \left(\theta\right) & = f_{\theta_c^{(1)}} \left(2 \arcsin \frac{\sin(\theta/2)}{\alpha^{(1)}\left( \Theta/N_l \right)}\right)  \frac{\cos(\theta/2)}{\sqrt{\left(\alpha^{(1)}\left( \Theta/N_l \right) \right)^2 - \sin^2(\theta/2)}}.
\end{split}
\end{equation}
\begin{proof}
    Since Lemma~\ref{lemma1} has provided the conditional coverage probability with a given single-hop distance, we can obtain the unconditional coverage probability of a single-hop by traversing the conditional coverage probability under different single-hop distances. The routing coverage probability is the product of the probability that each hop is covered. Following the Leibniz integral rule, the PDFs of central angle distributions are obtained by taking the derivative of CDFs.
\end{proof}
\end{theorem}

According to the definition, there may be cases where the SNR reaches the coverage threshold but the link is blocked by the Earth. Therefore, we denote availability and coverage probability $P^{A,C}_{\rm rout}$ as the probability that every hop is available, and every hop's received SNR is greater than the coverage threshold. $P^{A,C}_{\rm rout}$ can be calculated by replacing upper limit $\pi$ of integrals in (\ref{PCrout}) to $2\arcsin\frac{\sqrt{R_s^2-R_{\oplus}^2}}{R_s}$. Note that $P^{A,C}_{\rm rout} \neq P^A \times P^C_{\rm cond}$, since a hop's availability and coverage are not independent. The below theorem provides the analytical expression of routing transmission latency.

\begin{theorem}\label{theorem3}
    The average routing transmission latency of route $\mathcal{M}_{|N_l^*}$ obtained by Algorithm~\ref{alg2} is
\begin{sequation}\label{theo3_1}
\begin{split}
   & \overline{T}_{\rm tx} = 2 \int_0^{\pi} \int_0^{A_0} \int_0^{\infty} \varpi f_{\theta_c^{(1)}}(\Xi) f_{W|\theta_d}(w) f_{\theta_d}(\theta_d) B^{-1} \log_2 \left(1+\rho^t \, G \left( \frac{\lambda}{8\pi R_s\sin\frac{\Xi}{2}} \right)^2 \sigma^{-2} w \right)^{-1} \mathrm{d}\theta_d \, \mathrm{d}w \, \mathrm{d}\Xi \\
    & + \left(N_l-2\right) \int_0^{\pi} \int_0^{A_0} \int_0^{\infty}   \varpi f_{\theta_c^{(2)}}(\Xi) f_{W|\theta_d}(w) f_{\theta_d}(\theta_d)  B^{-1} \log_2 \left(1+\rho^t \, G \left( \frac{\lambda}{8\pi R_s\sin\frac{\Xi}{2}} \right)^2 \sigma^{-2} w \right)^{-1}  \mathrm{d}\theta_d \, \mathrm{d}w \, \mathrm{d}\Xi,
\end{split}
\end{sequation}
where $f_{\theta_c^{(1)}}(\Xi)$, $f_{\theta_c^{(2)}}(\Xi)$, $f_{W|\theta_d}(w)$ and $f_{\theta_d}(\theta_d)$ are defined in (\ref{fthetaPDF1}), (\ref{fthetaPDF2}), (\ref{pointing_error}) and (\ref{fthetad}), respectively. When communication is ultra-reliable (${\rm SNR} \gg 1$), an approximation for the routing transmission latency can be written as
\begin{equation}\label{approx1}
\begin{split}
    \widetilde{T}_{\rm tx} & = 2 \int_0^{\pi} f_{\theta_c^{(1)}}(\theta) T_{{\rm tx},1}^* \left(2R_s\sin\frac{\theta}{2}\right) \mathrm{d}\theta  + \left(N_l-2\right)\int_0^{\pi} f_{\theta_c^{(2)}}(\theta) T_{{\rm tx},1}^* \left(2R_s\sin\frac{\theta}{2}\right) \mathrm{d}\theta,
\end{split}
\end{equation}
where $T_{{\rm tx},1}^* \left(2R_s\sin\frac{\theta}{2}\right)$ is given in (\ref{T_tx}).
\begin{proof}
    See Appendix~\ref{app:theorem3}. 
\end{proof}
\end{theorem}

Next, some remarks on the extensibility of the above theorems are discussed. Firstly, these theorems are applicable to analyzing routing schemes with an arbitrary number of hops. Therefore, we can design other criteria to determine the number of hops' optimality, and other algorithms besides Algorithm~\ref{alg1} to find the optimal number of hops. In fact, these are what we will continue to discuss in the next section. Secondly, the analytical framework in this article applies to any distributions of $\theta_d$. For a given PDF of $\theta_d$, the coverage probability and latency can be derived in the same way by substituting this PDF in the proof of formulas (\ref{appA-3}) and (\ref{appB-1}).

\section{Optimization Through Mutual Relations (Method II)}\label{section4}
In Sec.~\ref{section3}, the $\varepsilon$-constraint method \cite{cui2017multi} is applied to transform a multi-objective optimization problem into a single-objective one. However, the above optimization method has the following three deficiencies: (i) Optimizing is not equivalent to constraining, and minimizing latency may degrade coverage performance. (ii) $\varepsilon$ needs to be carefully designed. A large $\varepsilon$ might cause (\ref{opt2-2}) to have no constraint effect, and a small $\varepsilon$ might result in no solution in the feasible region, as shown in Proposition~\ref{prop2}. (iii) 
When the link does not reach the coverage threshold and is interrupted, the data rate should be $0$. However, since the first method studies coverage probability and data rate separately, this issue can not be addressed in $\mathscr{P}_2$. Therefore, a new method is proposed to combine these three objectives into one objective.

\begin{table*}[]
\centering
\caption{Simulation Parameters \cite{al2021modeling,ata2022performance}.}
\label{table1}
\resizebox{\linewidth}{!}{ 
\renewcommand{\arraystretch}{1.1}
\begin{tabular}{|c|c|c||c|c|c|}
\hline
Notation     & Meaning                         & Default Value     & Notation   & Meaning    & Default Value         \\ \hline \hline
$R_{\oplus}$ & Radius of the Earth               & $6371$~km & $\varpi$   & Packet size & $10$~Mbit   \\ \hline
$R_{S}$         & {\scriptsize{Radius of the sphere where satellites are located}} & $7371$~km            & $c$ & Speed of light & $3 \times 10^8$~m/s \\ \hline
$\eta_s$, $A_0$ & Parameters of the pointing error                  & $1.00526$, $0.01979$ & $G$ & Antenna gain   & $160$~dBi           \\ \hline
$\varsigma$  & Variance of Rayleigh distribution & $15$~mrad & $\lambda$  & Wavelength   & $1550$~nm     \\ \hline
$\Theta$     & {\scriptsize{Central angle between the transmitter and receiver}}    & $\pi/2$    & $B$        & Bandwidth    & $20$~MHz     \\ \hline
$\rho_{t}$   & Transmit power of satellite       & $15$~dBW  & $\sigma^2$ & Noise power  & $10^{-10}$~mW \\ \hline
$\varepsilon$   & Tolerable probability of interruption       & 0.1  & $\gamma$ & Coverage threshold  & $0$~dB \\ \hline
\end{tabular}
}
\end{table*}

\subsection{ARQ Latency}
In this subsection, We introduce the ARQ protocol to combine the three objectives into a new optimization objective, based on their mutual relation. If the $i^{th}$ hop is available, and its received SNR is greater than the coverage threshold, the $i^{th}$ hop is "successfully" transmitted. The ARQ protocol requires the transmitter and receiver from the same hop to follow these communication principles: (i) the receiver should respond to the transmitter with "successful" when a packet has been received, and (ii) the transmitter sends the same packet repeatedly without waiting until the response "successful" is received. 

\par
"Successful" and "unsuccessful" transmission can be regarded as opposite events in a Bernoulli trial. For the $i^{th}$ hop, the number of trials required for a successful transmission satisfies the geometric distribution with $\frac{1}{P_i^{A,C}}$ as the mean value, where $P_i^{A,C}$ is the probability that the $i^{th}$ hop is available, and its received SNR is greater than the coverage threshold. As a result, transmission latency is weighted by $\frac{1}{P_i^{A,C}}$, and the original optimization problem $\mathscr{P}_1$ can be rewritten as 
\begin{equation}
\begin{split}
    \mathscr{P}_3: & \underset{\mathcal{M}_{|N_l}}{\mathrm{minimize}}  \ \underset{N_l}{\mathrm{minimize}}  \ \ T_{\rm ARQ} =  \sum_{i=1}^{N_l} \frac{1}{P_i^{A,C}} \left( \frac{\varpi }{B \log_{2}\left(1+{\rm{SNR}}_i \right)} \right),
\end{split}
\end{equation}
where $T_{\rm ARQ}$ is called routing transmission latency with ARQ protocol, which is abbreviated as ARQ latency.

\subsection{Metric Analysis and Optimization}
In this section, we discuss the discrete solution of $\mathscr{P}_3$ under the BPP model. The solving steps of $\mathscr{P}_3$ are similar to those of $\mathscr{P}_2$. Algorithm~\ref{alg2} is also applicable to the solution of $\mathscr{P}_3$, but Algorithm~\ref{alg1} requires to be changed in response to the new optimization objective. Considering the close relationship between availability, coverage probability, and latency, it is no longer suitable to optimize and analyze these three metrics separately. For the sake of joint optimization, we first give the analytic expression for the average value of $T_{\rm ARQ}$.
\begin{theorem}\label{theorem4}
    The average routing transmission latency with ARQ protocol of route $\mathcal{M}_{|N_l^*}$ obtained by Algorithm~\ref{alg2} is 
\begin{equation}\label{theo4_1}
\begin{split}
    & \overline{T}_{\rm ARQ} = 2  \int_0^{A_0} \int_0^{\infty} \int_0^{\theta_{\max}} \frac{\varpi f_{\theta_c^{(1)}}(\Xi) f_{W|\theta_d}(w) f_{\theta_d}(\theta_d) }{ P^C_{\rm cond} \left(2R_s \sin\frac{\Xi}{2}\right) B \log_2 \left(1+\rho^t \, G \left( \frac{\lambda}{8\pi R_s\sin\frac{\Xi}{2}} \right)^2 \sigma^{-2} w \right) }  \mathrm{d}\Xi \, \mathrm{d}\theta_d \, \mathrm{d}w \\
    & + \left(N_l-2\right) \int_0^{A_0} \int_0^{\infty} \int_0^{\theta_{\max}}  \frac{\varpi f_{\theta_c^{(2)}}(\Xi) f_{W|\theta_d}(w) f_{\theta_d}(\theta_d) }{ P^C_{\rm cond} \left(2R_s \sin\frac{\Xi}{2}\right) B \log_2 \left(1+\rho^t \, G \left( \frac{\lambda}{8\pi R_s\sin\frac{\Xi}{2}} \right)^2 \sigma^{-2} w \right) }  \mathrm{d}\Xi \, \mathrm{d}\theta_d \, \mathrm{d}w,
\end{split}
\end{equation}
where $f_{\theta_c^{(1)}}(\Xi)$, $f_{\theta_c^{(2)}}(\Xi)$, $f_{W|\theta_d}(w)$, $f_{\theta_d}(\theta_d)$ and $P^C_{\rm cond} \left(2R_s \sin\frac{\Xi}{2}\right)$ are defined in (\ref{fthetaPDF1}), (\ref{fthetaPDF2}), (\ref{pointing_error}), (\ref{fthetad}) and (\ref{PCcond}) respectively. $\theta_{\max} = 2\arcsin\left({\sqrt{R_s^2-R_{\oplus}^2}}/{R_s}\right)$ is the maximum central angle for an available hop. When communication is ultra-reliable (${\rm SNR} \gg 1$), an approximation for the routing transmission latency can be written as
\begin{sequation}\label{approx2}
\begin{split}
    \widetilde{T}_{\rm ARQ}  = 2 \int_0^{\theta_{\max}} f_{\theta_c^{(1)}}(\theta) \frac{T_{{\rm tx},1}^* \left(2R_s\sin\frac{\theta}{2}\right)}{P^C_{\rm cond} \left(2R_s \sin\frac{\theta}{2}\right)}  \mathrm{d}\theta 
     + \left(N_l-2\right) \int_0^{\theta_{\max}} f_{\theta_c^{(2)}}(\theta) \frac{T_{{\rm tx},1}^* \left(2R_s\sin\frac{\theta}{2}\right)}{P^C_{\rm cond} \left(2R_s \sin\frac{\theta}{2}\right)} \mathrm{d}\theta,
\end{split}
\end{sequation}
where $T_{{\rm tx},1}^* \left(2R_s\sin\frac{\theta}{2}\right)$ is given in (\ref{T_tx}).
\begin{proof}
    The proof of Theorem~\ref{theorem4} is similar to that of Theorem~\ref{theorem3}, therefore omitted here.
\end{proof} 
\end{theorem}

Similarly, we can apply the exhaustive search to obtain the optimal number of Hops. Since we lack a $\varepsilon$-constraint, we can introduce the following qualitative criteria to set an upper limit $N_{\max}$ for the exhaustive search. (i) When $n>N_{\max}$,  coverage probability, average routing transmission latency, and average routing transmission latency with ARQ protocol all decrease with $n$. (ii) When $n>N_{\max}$, the probability that two adjacent ideal satellite relay locations are associated with the same satellite cannot be ignored. (iii) $N_{\max} < \frac{\ln (1-\varepsilon)}{\ln(1-\varsigma^2)}$ is satisfied for a small $\varepsilon$. 
\par
Finally, based on the above analysis, we can conclude that the computational complexity of solving $\mathscr{P}_3$ in the discrete scenario is the same as that of $\mathscr{P}_2$.


\section{Numerical Results}\label{section5}
In this section, we demonstrate how the analytical results are applied for performance estimation through numerical results. Unless otherwise specified, the parameters in Sec.~\ref{section5} will be set to their default values from Table~\ref{table1}.

\subsection{Verification of Assumptions}\label{sec5-1}
This subsection investigates the reliability and latency performance of three deterministic constellations and validates the assumptions stated earlier. The constellation altitudes and the number of satellites are obtained from public data \cite{robert2020small}, while the remaining results in Table~\ref{table2} are calculated through Monte Carlo simulations and analytical results derived from theorems. The transmitter and receiver are assumed to be at the farthest ends, i.e., $\Theta=\pi/2$.

\par
By comparing the fourth and fifth rows of Table~\ref{table2}, it can be observed that the interference power of satellite communication is significantly smaller than the noise power. Therefore, the impact of interference power on SNR-related metrics can be ignored. Monte Carlo simulation is performed to calculate an upper bound for the interference of a typical satellite. We assume that the typical satellite associates with a satellite within its LoS range. The remaining satellites within the LoS range are associated with other satellites, excluding the typical satellite \cite{lou2023haps}. Each satellite aligns its transmission towards its associated satellite using a Gaussian antenna pattern \cite{gagliardi2012satellite}, resulting in bidirectional beam alignment between each pair of associated satellites.

\begin{table}[t]
\centering
\caption{Performances for Deterministic Constellations.}
\label{table2}
\renewcommand{\arraystretch}{1.2}
\begin{tabular}{|c|ccc|}
\hline
Constellation                        & \multicolumn{1}{c|}{Starlink} & \multicolumn{1}{c|}{Kuiper} & OneWeb \\ \hline
Number   of satellites               & \multicolumn{1}{c|}{11927}    & \multicolumn{1}{c|}{3236}   & 648    \\ \hline
Altitude   [km]                      & \multicolumn{1}{c|}{550}      & \multicolumn{1}{c|}{610}    & 1200   \\ \hline
Interference  $\times 10^{-16}$ [mw]                 & \multicolumn{1}{c|}{$27.81$} & \multicolumn{1}{c|}{$5.359$} & $0.908$ \\ \hline
Noise [mw]      & \multicolumn{3}{c|}{$10^{-10}$}    \\ \hline


Optimal number of hops (method I)   & \multicolumn{1}{c|}{5}   & \multicolumn{1}{c|}{5} & 6 \\ \hline
Optimal number of hops (method II)   & \multicolumn{1}{c|}{5}   & \multicolumn{1}{c|}{5} & 6 \\ \hline
Availability (Simu. \& Analy.)   & \multicolumn{1}{c|}{1.000}   & \multicolumn{1}{c|}{1.000} & 1.000 \\ \hline
Coverage   probability (Simu. \& Analy.)    & \multicolumn{1}{c|}{0.909}   & \multicolumn{1}{c|}{0.908} & 0.907 \\ \hline
Transmission   latency   [s]  (Simu. \& Analy.)  & \multicolumn{1}{c|}{0.642}   & \multicolumn{1}{c|}{0.645} & 0.741 \\ \hline
Approx.   transmission latency [s] (Analy.) & \multicolumn{1}{c|}{0.628}       & \multicolumn{1}{c|}{0.632}       & 0.729          \\ \hline
Propagation  latency [s]    (Simu.)   & \multicolumn{1}{c|}{0.071}   & \multicolumn{1}{c|}{0.072} & 0.079 \\ \hline
ARQ latency [s] (Simu. \& Analy.)  & \multicolumn{1}{c|}{0.654}   & \multicolumn{1}{c|}{0.658} & 0.754 \\ \hline
Approx.   ARQ latency   [s] (Analy.)  & \multicolumn{1}{c|}{0.641  }   & \multicolumn{1}{c|}{0.645} & 0.741 \\ \hline
\end{tabular}
\end{table}

\par
Optimization method I and II yield the same routing schemes for the three constellations since the optimal number of hops obtained by the two methods are consistent in Table~\ref{table2}. 
According to the proposed routing scheme, all constellations exhibit high routing coverage probabilities (around $0.9$) and single-hop coverage probabilities (around $0.98$). Routes are always available for these constellations. The total transmission latency is more than $0.6$ second, which is significantly longer than the propagation latency (less than $80$ms). The estimated approximate latency mentioned in the theorems is accurate compared to the exact latency obtained through simulation, and the deviation is only between $1.70\%$ and  $4.53\%$.

\subsection{Constellation Design} 
\begin{figure}[htbp]
\begin{minipage}[t]{0.49\linewidth}
\centering
\includegraphics[width=0.98\linewidth]{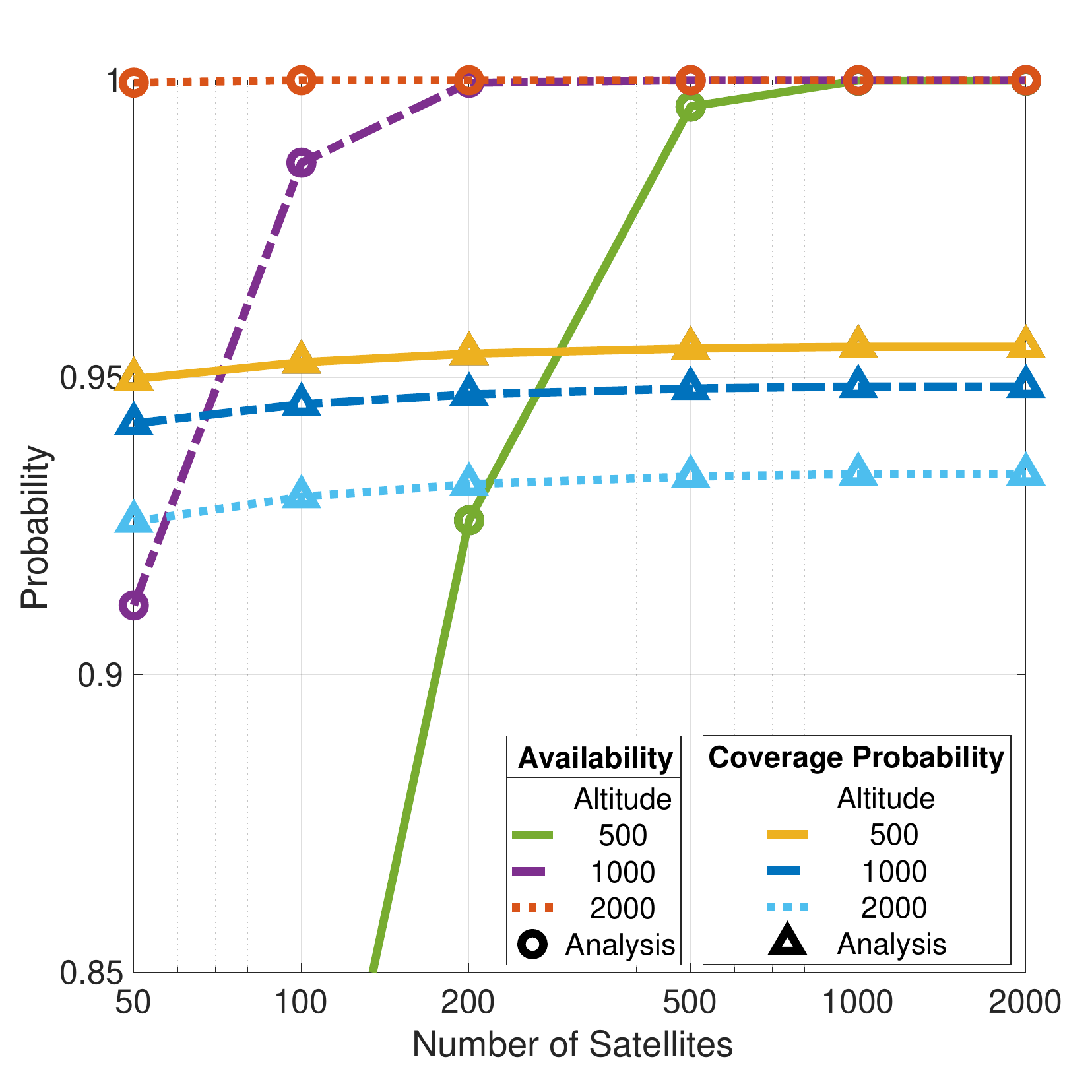}
\caption{Availability and coverage probability under different constellation configurations.}
\label{figure2}
\end{minipage}
\hfill
\begin{minipage}[t]{0.49\linewidth}
\centering
\includegraphics[width=0.98\linewidth]{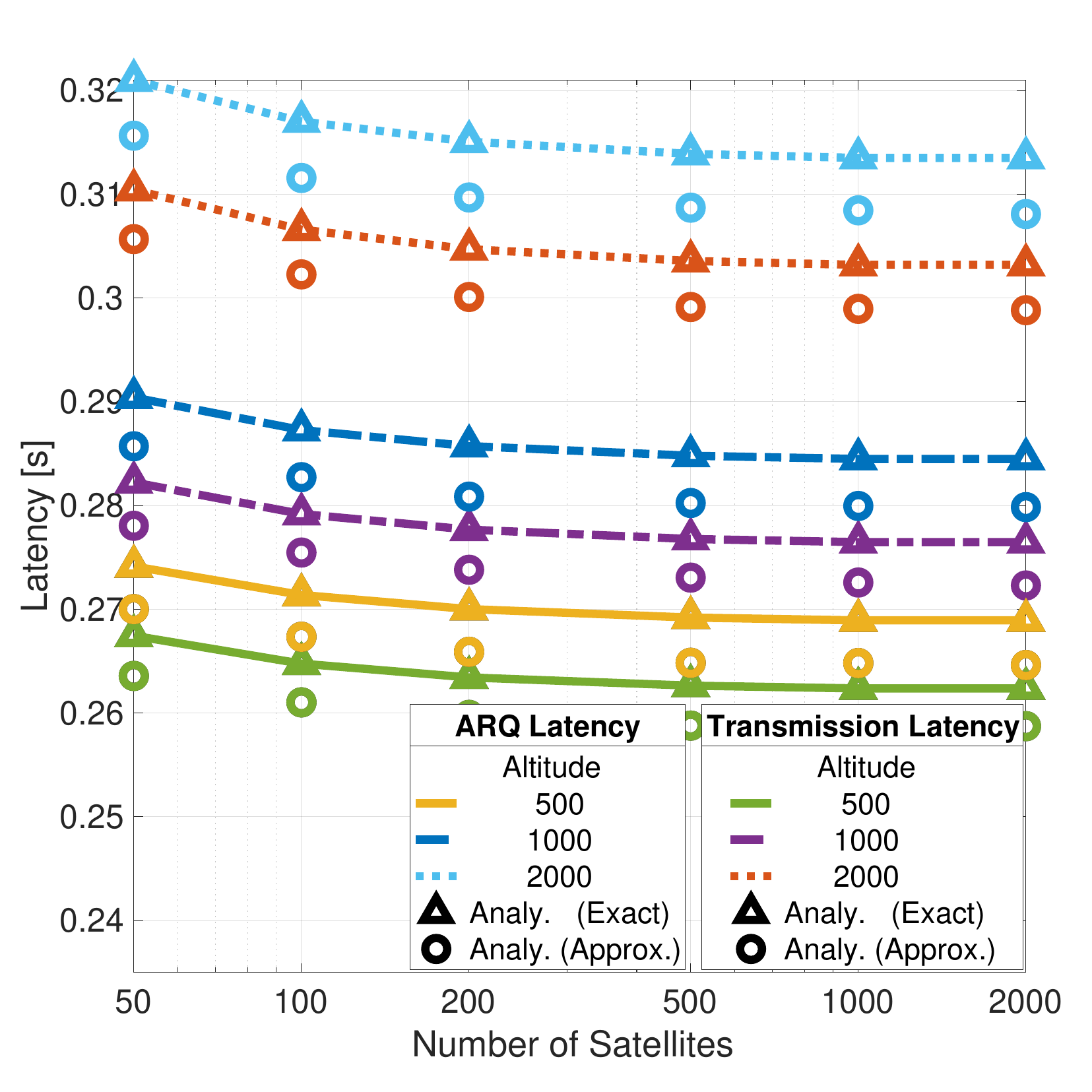}
\caption{Latency under different constellation configurations.}
\label{figure3}
\end{minipage}
\hfill
\end{figure}

This subsection aims to show insights into the design of satellite constellations. 
Fig.~\ref{figure2} indicates that, in the case of an abundant number of satellites (e.g., exceeding the default value of $1000$), the availability can approach $1$. The higher the deployment altitude of the satellites, the fewer satellites are required to achieve availability close to $1$. In contrast to availability, the routing coverage probability is negatively correlated with satellite altitude, given the same number of satellites. The impact of satellites' altitude on coverage probability is more significant than that of their number.

\par
Since both coverage probability and latency are SNR-based metrics, the conclusions of latency are similar to those of coverage probability. In Fig.~\ref{figure3}, satellite altitude also has a greater impact on latency relative to the number of satellites. A constellation with a lower satellite altitude and more satellites has a smaller latency. Overall, the ARQ latency obtained from method II in (\ref{approx2}) is larger than transmission latency obtained from method I in (\ref{approx1}). This is because when the link's SNR does not reach the coverage threshold, the transmission data rate in the calculation of ARQ latency is considered as $0$, while that of the transmission latency is not set to $0$. Under the same number of satellites and constellation altitude, the ARQ latency is approximately $3\%$ longer than the transmission latency. Both ARQ latency and transmission latency provide tight lower bounds for latency. As mentioned, when the routing strategy achieves lower latency and higher reliability, the approximate results will be more accurate. 

\subsection{Strategy Comparison}
We showcase the advantages of the proposed algorithm over existing tractable algorithms in terms of transmission latency and ARQ latency. In labels of Fig.~\ref{figure4} and Fig.~\ref{figure5}, the ideal solution refers to the routing path in an ideal scenario proposed in Sec.~\ref{sec3-1}, which can serve as a lower bound for latency that cannot be achieved. The remaining three routing strategies, which have been verified to be tractable through the SG framework, are used as benchmarks:

\begin{figure}[htbp]
\begin{minipage}[t]{0.49\linewidth}
\centering
\includegraphics[width=0.98\linewidth]{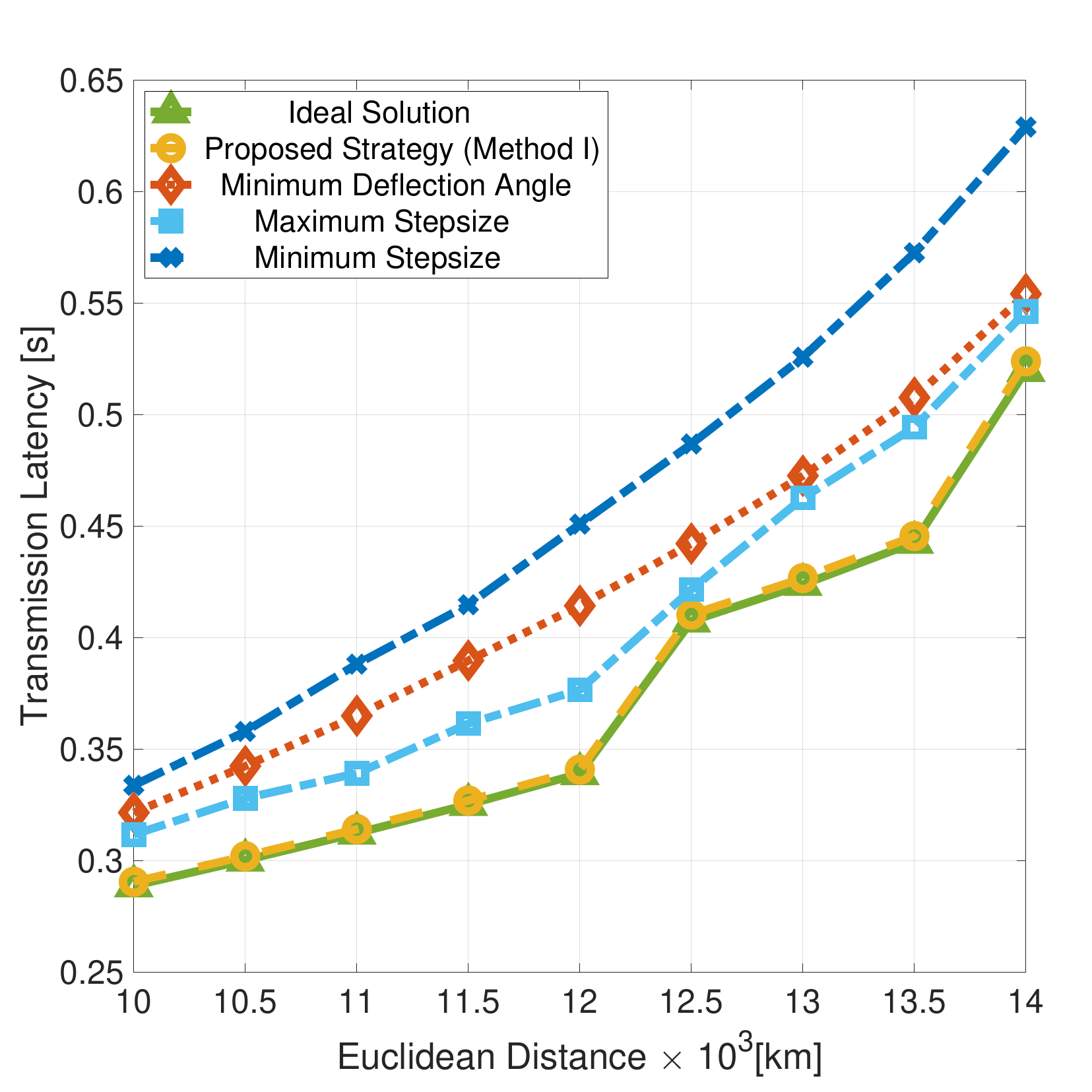}
\caption{Comparison of routing strategies at different communication distances.}
\label{figure4}
\end{minipage}
\hfill
\begin{minipage}[t]{0.49\linewidth}
\centering
\includegraphics[width=0.98\linewidth]{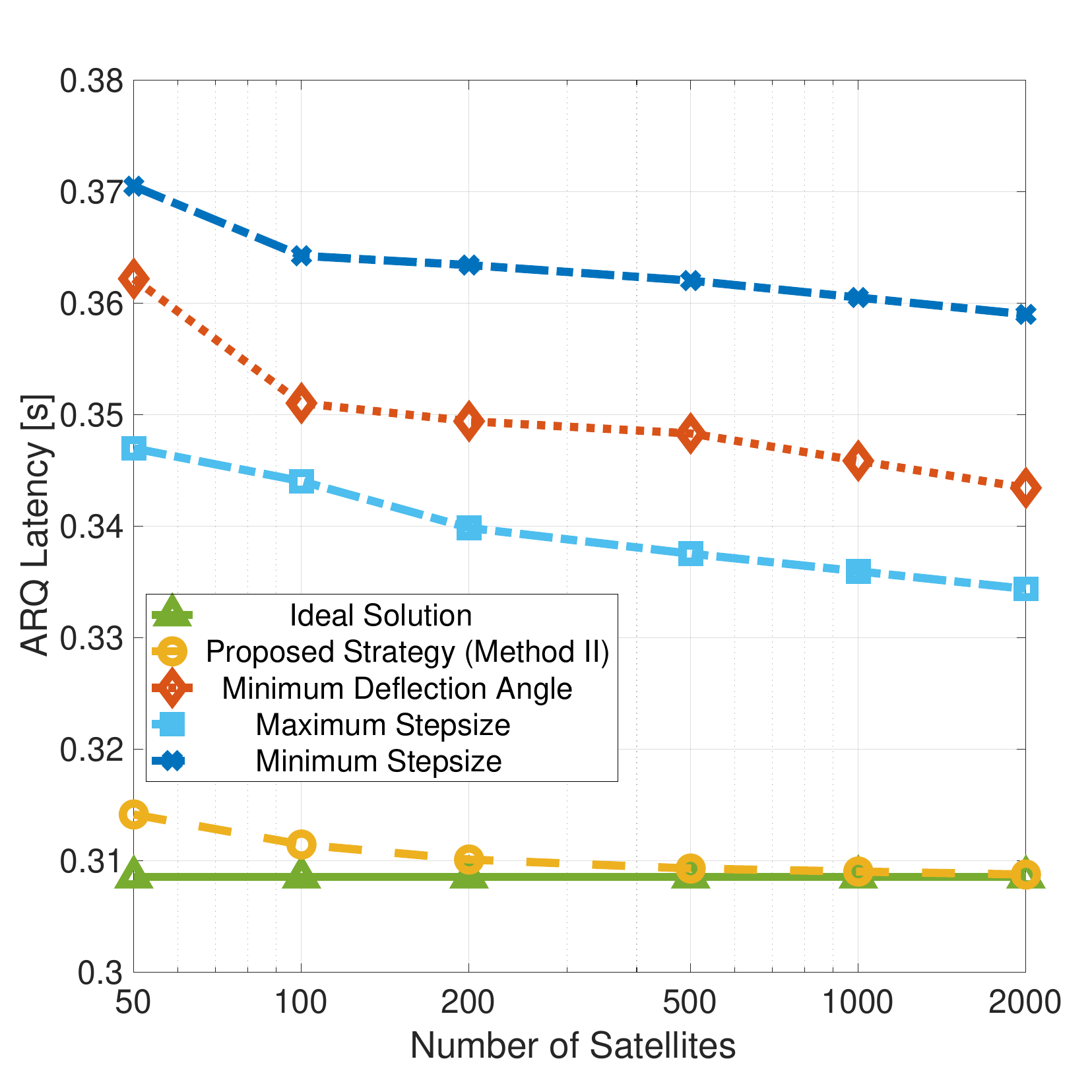}
\caption{Comparison of routing strategies with different numbers of satellites.}
\label{figure5}
\end{minipage}
\end{figure}

\begin{itemize}
    \item Minimum deflection angle: the satellite within the communication range \footnote{The communication range in this article is determined based on the constraint specified in (\ref{opt4-2}).} and with the least deflection from the shortest inferior arc is selected as the next hop \cite{wang2022stochastic}.
    \item Maximum stepsize: the satellite within the communication range and closest to the receiver is selected as the next hop \cite{lou2023coverage}.
    \item Minimum stepsize: the satellite closest to the previous hop within a given direction angle ($\pi/6$ in this article) is selected as the next hop \cite{haenggi2005routing}.
\end{itemize}

\par
Fig.~\ref{figure4} investigates the influence of the Euclidean distance between the transmitter and receiver on routing transmission latency when the number of satellites is fixed at $1000$. The routing transmission latency corresponding to the proposed strategy can approach the lower bound of latency in the ideal scenario. Compared to the other three benchmark strategies, the proposed strategy exhibits significantly smaller latency, especially for a shorter communication distance. Fig.~\ref{figure5} demonstrates that the advantages of the proposed routing strategy are also reflected in the ARQ latency. In addition, a straightforward conclusion from Fig.~\ref{figure5} is that the gap between the ARQ latency obtained by the proposed strategy and the ARQ latency in the ideal scenario decreases as the number of satellites increases.

\subsection{Extension to Satellite-Terrestrial Communication}
\begin{figure}[htbp]
\begin{minipage}[t]{0.49\linewidth}
\centering
\includegraphics[width=0.98\linewidth]{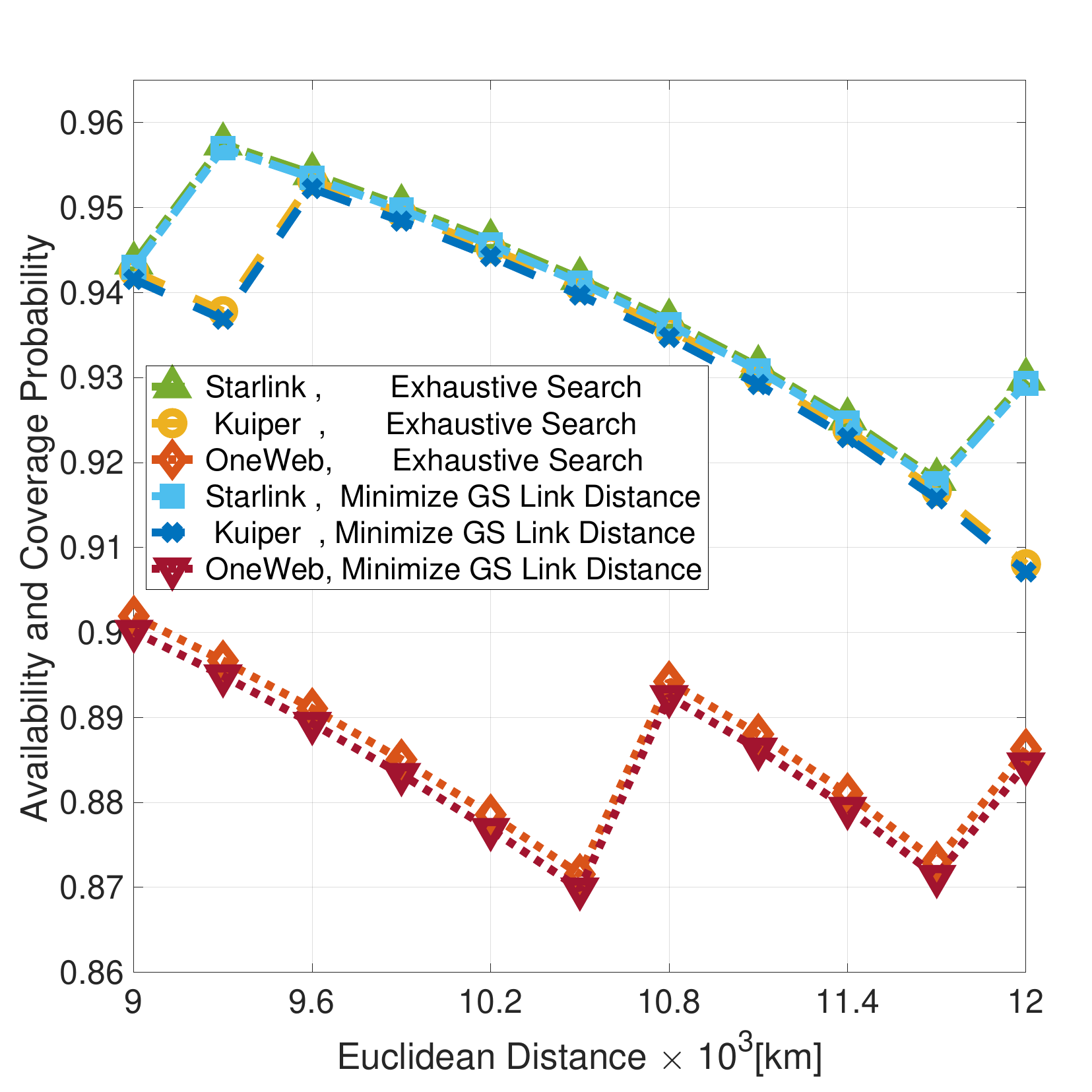}
\caption{Reliability performance of deterministic constellations.}
\label{figure6}
\end{minipage}
\hfill
\begin{minipage}[t]{0.49\linewidth}
\centering
\includegraphics[width=0.98\linewidth]{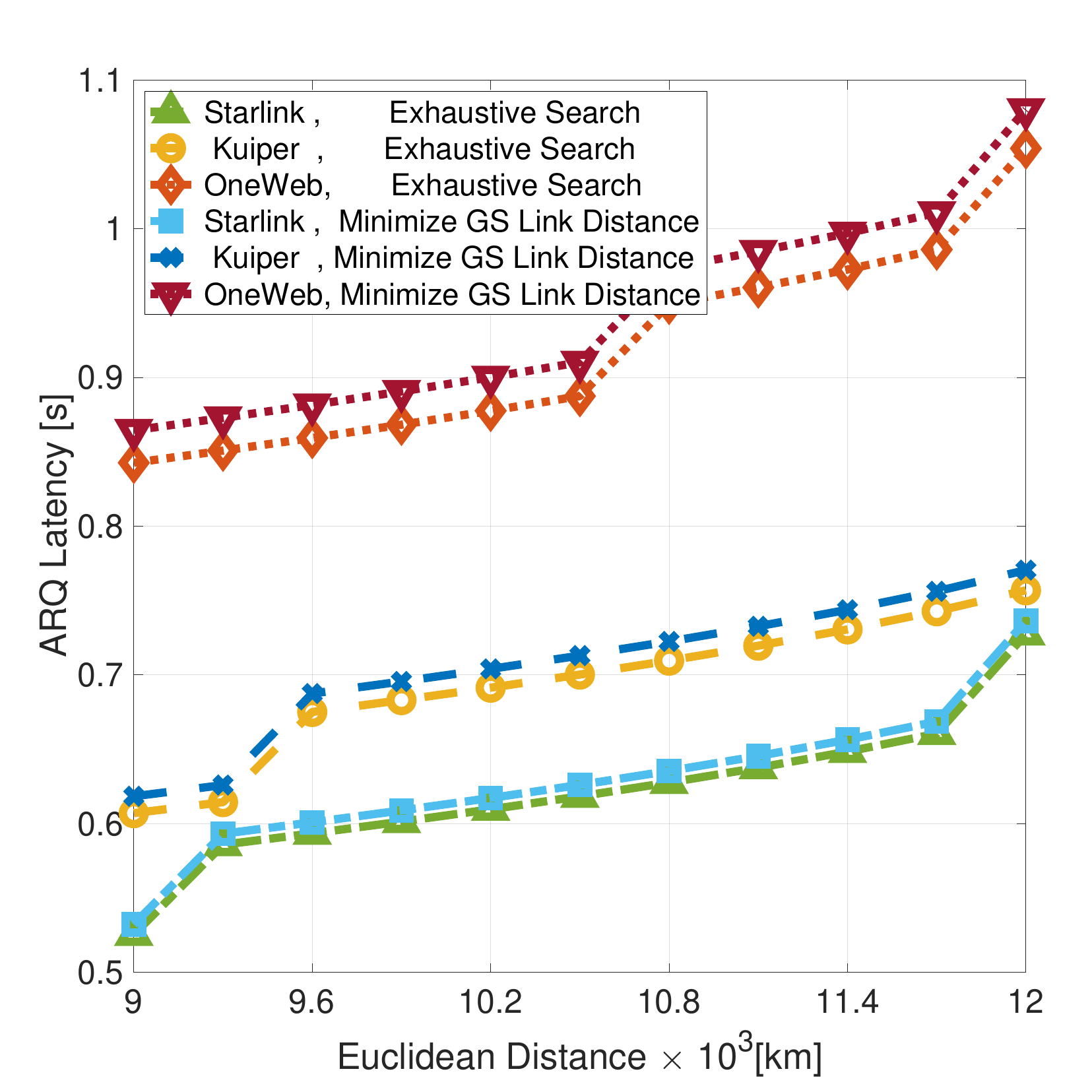}
\caption{Latency performance of deterministic constellations.}
\label{figure7}
\end{minipage}
\end{figure}

Considering that satellite communication requests are typically initiated and received on the ground, this subsection discusses the scenario where communication is initiated from a ground transmitter, transmitted through multiple satellite relays, and returned to a ground receiver. In Fig.~\ref{figure6} and Fig.~\ref{figure7}, the parameters of three deterministic constellations are given in Table~\ref{table1}. The label "exhaustive search" refers to traversing through possible combinations of relay positions and selecting the combination with the minimum latency. Although the distances between the first hop, intermediate hops, and the last hop differ, considering that the total communication distance is fixed, the "exhaustive search" method has only two degrees of freedom and has quadratic complexity. The label "minimize GS link distance" refers to minimizing the distances between the first hop and the last hop while intermediate hops are evenly distributed.

\par
Under the same constellation configuration, the two routing strategies exhibit tiny differences in terms of reliability and latency performance. This conclusion indicates that the analytical results of this article can be extended to the satellite-terrestrial communication scenario by simply combining them with the findings in \cite{ok-1}, which analyzed the performance under the nearest satellite association of ground stations. Furthermore, instead of smoothly varying with the communication distance, there are discontinuous jumps in the performance of the availability and coverage probability, and the ARQ latency, resembling stepped lines. The appearance of such curves in Fig.~\ref{figure6} and Fig.~\ref{figure7} (as well as Fig.~\ref{figure4}) is due to the discrete nature of the changes in the number of hops.

\section{Further Discussions}
\subsection{Deterministic Channel Parameters}
The SG-based approach is less system-dependent compared to deterministic methods, and therefore has stronger scalability. In other words, the analytical expressions provided in this article are functions of system and channel parameters. Consequently, when the numerical values of the parameters change, the SG-based analytical framework remains applicable, and the derived analytical expressions can still achieve a low-complexity and accurate mapping from parameters to performance estimates. We use the example of an ISL channel to illustrate why the scalability of this advantage is highly valued. The ISL is influenced by a multitude of factors such as antenna directionality, frequency selection, and satellite operational status. Many of these factors undergo changes with different constellations, and some are even unpredictable. For instance, operational status-related factors such as satellite aging may cause unpredictable jitter amplitudes.

\par
However, the advantage of scalability is achieved at the cost of performance. When more information about the channel state is known, we can use the SG-based analytical framework as a basis for further strategy design and achieve better reliability and latency performance. The width of the beam can be adjusted according to the satellite's jitter condition. For satellites experiencing severe pointing error, it is advisable to appropriately sacrifice beam gain and select a beam with a wider main lobe width. Satellite selection can be adjusted based on the satellite's payload situation to achieve load balancing. Furthermore, for a link with higher path loss, it is advisable to increase the transmission power appropriately. For scenarios where these channel parameters are determined, the SG-based analytical framework can provide initial and coarse-grained predictions.

\subsection{Deterministic Topology}
Within the SG framework, the algorithm proposed in this article can offer a low-complexity relay selection strategy when satellite positions are unknown. On the other hand, fixed-topology satellite routing algorithms not only rely on accurate satellite ephemeris but also require real-time updates of constellation status, such as load information. Furthermore, the complexity of fixed-topology algorithms is relatively higher, and the performance of the algorithm at the system level is difficult to analyze. While the requirements for implementing such algorithms are demanding, with sufficient known information, abundant time, and computational support, fixed-topology algorithms undoubtedly have the potential to outperform SG-based routing algorithms. 

\par
Therefore, investigating the impact of additional information, such as satellite status, on routing performance is a promising research direction. For example, give that the topology and load information are available, we can design load-balanced routing and apply the algorithm proposed in this article as a benchmark for comparison. Additionally, by implementing the SG-based algorithm, the initial selection of relay satellites for fixed topology-based routing can be expedited, thereby accelerating algorithm convergence.

\subsection{Doppler Effect}
In the channel model, we do not consider the potential impact of the Doppler effect on communication quality. However, due to the rapid relative motion between satellites, the Doppler effect cannot be ignored. Fortunately, satellite communication can mitigate the effect by conducting bit synchronization corrections and expanding the range of the receiving frequency \cite{ziedan2003bit}. Compared to terrestrial short-distance communication, the transmission rate of ISL is significantly lower, making the bit synchronization less difficult as well \cite{kaddoum2012design}. Furthermore, the bandwidth is not the primary constraint for satellite communication. Due to the directional beam alignment in ISL, expanding the range of receiving frequencies does not introduce excessive co-frequency interference.

\section{Conclusion}
To meet the requirements of high reliability and low latency, we have designed a routing strategy for LEO satellite routing and analyzed its performance. The current deployment approach of densely populated low-altitude constellations in use is advantageous for improving communication reliability and reducing latency. In terms of latency, the proposed routing strategy demonstrated significant advantages compared to existing strategies and approached the ideal lower bound of latency. This strategy, along with its corresponding analytical framework, can be easily extended to routing with satellite-to-ground links.

\appendices
\section{Proof of Proposition~\ref{prop1}}\label{app:prop1}
The proof of proposition~\ref{prop1} is divided into the following four parts: (i) the existence of feasible region, (ii) potential optimal relay position distribution, (iii) the uniqueness of the solution, and (iv) the corresponding minimum latency.

\par
There are an infinite number of planes passing through the transmitting and receiving satellites, and there are an infinite number of arcs intersected by those planes with the sphere where satellites are located. The shortest route between the transmitting satellite and the receiving satellite should be along with the shortest inferior arc corresponding to the string connected by the two satellites. For a given arc, the relationship between the arc length $l_{\rm arc}$ and the radius of the circle $r$ corresponding to the arc is
\begin{equation}
    l_{\rm arc} = 2r \arcsin\frac{l_{\rm str}}{2r},
\end{equation}
where $l_{\rm str}=2R_s \sin\frac{\Theta}{2}$ is the Euclidean distance between the transmitter and receiver. When $2r>l_{\rm str}$, apply Taylor series expansion to $\arcsin\frac{\Theta}{2}$, we have
\begin{equation}
    l_{\rm arc} = l_{\rm str} + \frac{l_{\rm str}^3}{6 (2r)^2} + \frac{3l_{\rm str}^5}{40 (2r)^4} + \cdots
\end{equation}
It is not hard to find $l_{\rm arc}$ decreases with $r$. The shortest route/inferior arc should be centered on the center of the Earth because the radius of the arc $r=R_s$ is the largest in this case. The azimuth angle of optimal positions of relay satellites located on the shortest inferior arc should be $0$, and positions of the $i^{th}$ relay satellite can be expressed as $\left(R_s,\psi_i,0\right)$. Considering that the length of a single-hop satisfies $l_i \leq 2\sqrt{R_s^2 - R_\oplus^2}$, for a route along the shortest inferior arc, the maximum central angle of each hop is $2\arccos\frac{R_\oplus}{R_s}$. Therefore, the following constraint needs to be satisfied to ensure the routing can be finished within $N_l$ hops:
\begin{equation}
    2 N_l \sqrt{R_s^2 - R_\oplus^2} \geq \Theta.
\end{equation}
Satisfying the above condition means that the feasible region exists. 

\par
In the ideal scenario, the problem of satellite subset selection can be equivalent to the optimization of each hop's distance. Given that the positions of relay satellites are located on the shortest inferior arc, $\mathscr{P}_{2-1}$ can be rewritten as,
\begin{subequations} 
	\begin{alignat}{2}
		\underset{N_l}{\mathrm{minimize}} \ \ \ &  \sum_{i=1}^{N_l} \frac{\varpi }{B \log_{2}\left(1+\mathbb{E}_W[{\rm{SNR}}_i] \right)}, \\
		\mathrm{subject \ to} \ \ & \sum_{i=1}^{N_l} \theta_i 
        = \sum_{i=1}^{N_l} 2\arcsin\frac{l_i}{2R_s} = \Theta.
	\end{alignat}
\end{subequations}
We introduce Lagrange multiplier $\mu$ to solve this constrained problem and let
\begin{equation}
\begin{split}
    \frac{\partial}{\partial l_i} \Bigg( \sum_{i=1}^{N_l} \frac{\varpi }{B \log_{2}\left(1+\rho^t \, G \left( \frac{\lambda}{4\pi l_i} \right)^2 \mathbb{E}[W] \sigma^{-2} \right)}  
    + \mu \left( \sum_{i=1}^{N_l} 2\arcsin\frac{l_i}{2R_s} - \Theta \right) \Bigg) = 0. 
\end{split}
\end{equation}
Then, the following differential equation exists for $\forall i$:
\begin{sequation}\label{appA-1}
    \frac{\partial}{\partial l_i}  \frac{\varpi }{B \log_{2}\left(1+\rho^t \, G \left( \frac{\lambda}{4\pi l_i} \right)^2 \mathbb{E}[W] \sigma^{-2} \right)} = \frac{\partial}{\partial l_i} 2 \mu \arcsin\frac{l_i}{2R_s}.
\end{sequation}
Therefore, there exists a $\mu$ satisfies the above equation set when $l_1=l_2= \cdots = l_{N_l}$, and the corresponding positions of the relay satellite are 
\begin{equation}
    \left\{ \left(R_s,\frac{\Theta}{N_l},0\right), \left(R_s,\frac{2\Theta}{N_l},0\right), \dots, \left(R_s,\frac{(N_l-1)\Theta}{N_l},0\right) \right\}.
\end{equation}

\par
Solving differential equations (\ref{appA-1}) is complicated, but we can discuss the uniqueness of the solution in special cases. When $\mathbb{E}_W [{\rm SNR}_i] \ll 1$, 
\begin{equation}\label{appA-2}
\begin{split}
    & \log_2 \left( 1 + \mathbb{E}_W [{\rm SNR}_i] \right) = \frac{1}{\ln 2} \ln \left( 1 + \mathbb{E}_W [{\rm SNR}_i] \right) \approx \frac{\mathbb{E}_W [{\rm SNR}_i]}{\ln 2}.
\end{split}
\end{equation}
Substitute (\ref{appA-2}) into (\ref{appA-1}), we get
\begin{equation}
    \frac{2 \ln 2 \varpi \sigma^2}{B \rho^t G \left(\frac{\lambda}{4\pi}\right)^2 \mathbb{E}[W]} l_i = \frac{2 \mu}{\sqrt{4R_s^2 - l_i^2}},
\end{equation}
\begin{equation}
    \mu = \frac{ \ln 2 \varpi \sigma^2}{B \rho^t G \left(\frac{\lambda}{4\pi}\right)^2 \mathbb{E}[W]} \sqrt{4R_s^4 - \left( 2 R_s^2 - l_i^2 \right)^2}, \ \ {\rm for \ } \forall i.
\end{equation}
Since $l_i^2 \leq 4 \left(R_s^2 - R_{\oplus}^2 \right) <  2 R_s^2 $ for LEO satellites, $\mu$ monotonously increases with $l_i$. Therefore, $l_1=l_2= \cdots = l_{N_l}$ is always true when $\mathbb{E}_W [{\rm SNR}_i] \ll 1$. Unfortunately, $\mu$ does not change monotonically with $l_i$ when $\mathbb{E}_W [{\rm SNR}_i] \gg 1$. After further research, the conclusions about the uniqueness of solutions are summarized as follows.
\par
(i) When communication is unreliable ($\mathbb{E}_W [{\rm SNR}_i] \ll 1$), the minimum latency can be reached when the satellites are distributed equally at the shortest inferior arc. 
(ii) When communication is reliable ($\mathbb{E}_W [{\rm SNR}_i] \gg 1$), the above satellite distribution can reach the local minimum latency, not necessarily the global minimum. (iii) When communication is unreliable ($\mathbb{E}_W [{\rm SNR}_i] \ll 1$), according to the above distribution, increasing the number of hops can always reduce the latency. (iv) When communication is reliable ($\mathbb{E}_W [{\rm SNR}_i] \gg 1$) and $N_l$ is larger than the optimal number of hops $N_l^*$, the above satellite distribution can not reach the global minimum latency.

\par
Finally, take the expectation to the small scale fading,
\begin{equation}\label{appA-3}
\begin{split}
    & \mathbb{E}[W]  = \int_0^{A_0} \int_0^{\infty} w f_{W|\theta_d}\left ( w \right ) f_{\theta_d}\left ( \theta_d \right ) \mathrm{d} \theta_d \mathrm{d} w = \int_0^{A_0}  \frac{\eta_s^2 w^{\eta_s^2 } }{A_0^{\eta_s^2}} \mathrm{d} w \int_0^{\infty} \cos\theta_d \frac{\theta_d}{\varsigma^2}\exp\left ( -\frac{\theta_d^2}{2\varsigma^2} \right ) \mathrm{d} \theta_d \\
    & \overset{(a)}{\approx} \frac{A_0 \eta_s^2}{1 + \eta_s^2} \left( 1 - \frac{1}{2} \int_0^{\infty} \frac{\theta_d^3}{\varsigma^2}\exp\left ( -\frac{\theta_d^2}{2\varsigma^2} \right ) \mathrm{d}\theta_d \right) \\ 
    & \overset{(b)}{=} \frac{A_0 \eta_s^2}{1 + \eta_s^2} \left( 1 - \int_0^{\infty} x\varsigma^2  \exp\left ( -x \right ) \mathrm{d}x \right) \overset{(c)}{=} \frac{A_0 \eta_s^2}{1 + \eta_s^2} \left( 1 - \varsigma^2 \right),
\end{split}
\end{equation}
where step (a) follows the second-order Taylor expansion of $\cos\theta_d\approx 1 - \frac{\theta_d^2}{2}$ ($\theta_d$ is generally a small value), step (b) is derived by the substitution of $x = \frac{\theta_d^2}{2 \varsigma^2}$, and step (c) follows the mean of the exponential distribution. The $i^{th}$ hop's local minimum latency under the potential optimal satellite relay distribution can be expressed as
\begin{sequation}
\begin{split}
    & T_{{\rm tx},1}^* \left(l_i\right) = \frac{\varpi}{B \log_2 \left(1+\rho^t \, G \left( \frac{\lambda}{4\pi l_i} \right)^2 \mathbb{E}[W] \sigma^{-2} \right) }  = \frac{\varpi}{B \log_2 \left(1+\rho^t \, G \left( \frac{\lambda}{4\pi l_i} \right)^2 \frac{A_0 \eta_s^2}{1 + \eta_s^2} \left( 1 - \varsigma^2 \right) \sigma^{-2} \right) }.
\end{split}
\end{sequation}

\section{Proof of Lemma~\ref{lemma1}}\label{app:lemma1}
According to the definition of conditional coverage probability, $P^C_{\rm cond} (l_i)$ can be expressed as follows: 
\begin{sequation}\label{appB-1}
\begin{split}
    & P^C_{\rm cond} (l_i)  = \mathbb{P} \left[ \rho^t \, G \left( \frac{\lambda}{4\pi l_i} \right)^2 W \sigma^{-2} \geq \gamma \Big| \, l_i \right]  = \mathbb{P} \left[ W \geq \frac{\gamma \sigma^2 \left( 4\pi l_i \right)^2}{\rho^t \, G \lambda^2} \bigg| \, l_i \right] \\
    & \overset{(a)}{=} \int_{\frac{\gamma \sigma^2 \left( 4\pi l_i \right)^2}{\rho^t \, G \lambda^2}}^{A_0} f_{W}\left ( w \right ) \mathrm{d}w  \overset{(b)}{=} \int_{\frac{\gamma \sigma^2 \left( 4\pi l_i \right)^2}{\rho^t \, G \lambda^2}}^{A_0} \int_0^{\infty} f_{W|\theta_d}\left ( w \right ) f_{\theta_d}\left ( \theta_d \right ) \mathrm{d}\theta_d \mathrm{d}w, \\
\end{split}
\end{sequation}
where step (a) follows the definition of \ac{CCDF}, and step (b) holds because the deviation angle $\theta_d$ is independent of pointing error $W$. Substitute $f_{W|\theta_d}\left ( w \right ) = \frac{\eta_s^2 w^{\eta_s^2-1 } \cos\theta_d}{A_0^{\eta_s^2}}$ and $f_{\theta_d}\left ( \theta_d \right )=\frac{\theta_d}{\varsigma^2}\exp\left ( -\frac{\theta_d^2}{2\varsigma^2} \right )$ into (\ref{appB-1}), we have
\begin{sequation}\label{appB-2}
\begin{split}
    &P^C_{\rm cond} (l_i) = \int_{\frac{\gamma \sigma^2 \left( 4\pi l_i \right)^2}{\rho^t \, G \lambda^2}}^{A_0} \int_0^{\infty} \frac{\eta_s^2 w^{\eta_s^2-1 } \cos\theta_d}{A_0^{\eta_s^2}}  \frac{\theta_d}{\varsigma^2}\exp\left ( -\frac{\theta_d^2}{2\varsigma^2} \right ) \mathrm{d}\theta_d \mathrm{d}w \\
    & = \frac{\eta_s^2}{A_0^{\eta_s^2} \varsigma^2} \int_{\frac{\gamma \sigma^2 \left( 4\pi l_i \right)^2}{\rho^t \, G \lambda^2}}^{A_0} w^{\eta_s^2-1 } \mathrm{d}w \int_0^{\infty} \theta_d \cos\theta_d  \exp\left ( -\frac{\theta_d^2}{2\varsigma^2} \right ) \mathrm{d}\theta_d  \overset{(c)}{\approx} \left( 1 - \left(\frac{\gamma \sigma^2 \left( 4\pi l_i \right)^2}{A_0 \rho^t \, G \lambda^2}\right)^{\eta_s^2} \right) \left( 1 - \varsigma^2 \right),
\end{split}
\end{sequation}
where step (c) follows the second-order Taylor expansion of $\cos\theta_d\approx 1 - \frac{\theta_d^2}{2}$.

\section{Proof of Lemma~\ref{lemma2}}\label{app:lemma2}
In this appendix, we will prove Lemma~\ref{lemma2} according to the following steps: (i) obtaining the neatest neighbor distribution, (ii) deriving the distance scaling factor of the first/last hop, and (iii) approximately estimating the distance scaling factor of a middle hop according to that of the first/last hop.

\par
According to Slivnyak's theorem \cite{feller1991introduction}, the distribution of a spherical homogeneous BPP is invariant with the rotation of the sphere where the BPP is located. To simplify the derivation of the distance scaling factor of the first or last hop, We assume that one reference point represents the transmitter or receiver located at $\left( R_s,\theta_i,0 \right)$. The other reference point is located at $\left( R_s,0,0 \right)$, thus $\theta_i$ is the central angle between two reference points. 

\par
Now, we derive the nearest neighbor distribution of the reference point at $\left( R_s,0,0 \right)$. The nearest neighbor's azimuth angle $\phi_{nn}$ is uniformly distributed between $0$ and $2\pi$. Denote $\mathcal{S}(\theta)$ as the spherical cap with a central angle as $2\theta$, whose rotation axis is the line connecting the reference point at $\left( R_s,0,0 \right)$ and the center of the Earth. Then, CDF of the nearest neighbor's polar angle $\psi_{nn}$ can be calculated as
\begin{equation}
\begin{split}
    & F_{\psi_{nn}}(\theta) = \mathbb{P}\left[ \psi_{nn} \leq \theta \right]  = 1 - \mathbb{P}\left[ {\mathcal{N}\left( {{\mathcal{S}(\theta)}} \right) = 0} \right] = 1 - \left( 1 - \frac{\mathcal{A}\left( \mathcal{S}(\theta) \right)}{\mathcal{A}\left( \mathcal{S}(\pi) \right)} \right)^{N_s} \\
    & = 1 - \left( 1 - \frac{ 2\pi R_s^2 (1-\cos\theta)}{4\pi R_s^2} \right)^{N_s} = 1 - \left( \frac{ 1 + \cos\theta }{2} \right)^{N_s},
\end{split}
\end{equation}
where $N_s$ is the number of satellites, $\mathcal{N}\left( \mathcal{S}(\theta) \right)$ counts the number of satellites in the spherical cap $\mathcal{S}(\theta)$, and $\mathcal{A}\left( \mathcal{S}(\theta) \right)$ is the area measure of $\mathcal{S}(\theta)$. Furthermore, the PDF of $\psi_{nn}$ is
\begin{equation}\label{fpsinn}
\begin{split}
    f_{\psi_{nn}}(\theta) = \frac{\mathrm{d}}{\mathrm{d}\theta} F_{\psi_{nn}}(\theta) = \frac{N_s \sin\theta}{2} \left( \frac{ 1 + \cos\theta }{2} \right)^{N_s-1}.
\end{split}
\end{equation}

\par
Next, we calculate the average distance from the reference point $\left( R_s,\theta_i,0 \right)$ to the nearest neighbor of $\left( R_s,0,0 \right)$ by traversing the location of this nearest neighbor. Denote the above average distance as $\overline{d^{(1)}}$, we have
\begin{equation}
    \overline{d^{(1)}} = \int_0^{2\pi} \int_0^{\pi} \frac{1}{2\pi} f_{\psi_{nn}}(\xi) d\left(\theta_i,0;\xi,\varphi \right) \mathrm{d}\xi \mathrm{d}\varphi,
\end{equation}
where $d\left(\theta_i,0;\xi,\varphi \right)$ is given in (\ref{d_12}). The distance scaling factor of the first or last hop is defined as the ratio of $\overline{d^{(1)}}$ to the distance of the first or last hop in the ideal scenario,
\begin{sequation}\label{appC-1}
\begin{split}
    \alpha^{(1)} \left(\theta_i\right) = \frac{\overline{d^{(1)}}}{2R_s \sin\frac{\theta_i}{2}} = \frac{N_s}{8\pi R_s \sin\frac{\theta_i}{2}} \int_0^{2\pi} \int_0^{\pi} \sin\xi \left( \frac{ 1 + \cos\xi }{2} \right)^{N_s-1} d\left(\theta_i,0;\xi,\varphi \right) \mathrm{d}\xi \mathrm{d}\varphi.
\end{split}
\end{sequation}

\par
Finally, we provide two approximate evaluations for the distance scaling factor of a middle hop. In the first approximation, we assume that one reference point's distance increment caused by matching the nearest neighbor is independent of the other one. Therefore, we get the additive evaluation: $\alpha^{(2)} \left(\theta_i\right) = 2\alpha^{(1)} \left(\theta_i\right) - 1$. In the second approximation, we assume two matching procedures bring increments one by one. From this perspective, $d\left(\theta_i,0;\xi,\varphi \right)$ in (\ref{appC-1}) is substituted by $\alpha^{(1)} \left(\theta_i\right) d\left(\theta_i,0;\xi,\varphi \right)$ after the first round's increment is in effect. By this way, we get the multiplicative evaluation: $\alpha^{(2)} \left(\theta_i\right) = \left(\alpha^{(1)} \left(\theta_i\right)\right)^2$.

\section{Proof of Theorem~\ref{theorem1}}\label{app:theorem1}
We start the proof with the derivation of CDF of type-I central angle distribution. Similar to Appendix~\ref{app:lemma2}, we assume one reference point represents the transmitter or receiver located at $\left( R_s,0,0 \right)$. The other reference point is located at $\left( R_s,\frac{\Theta}{N_l},0 \right)$, without loss of generality. Given that the central angle between the nearest neighbor and $\left( R_s,0,0 \right)$ is less than $\Xi$, we traverse potential positions where the nearest neighbor may occur, the PDF of the nearest neighbor's polar angle $\psi_{nn}$ and azimuth angle $\phi_{nn}$ is
\begin{equation}
    f_{\psi_{nn}, \phi_{nn}}(\theta) = \frac{f_{\psi_{nn}}(\theta)}{2\pi R_s^2 \sin{\theta}} = \frac{N_s}{4\pi R_s^2} \left( \frac{ 1 + \cos\theta }{2} \right)^{N_s-1}
\end{equation}
where ${f_{\psi_{nn}}(\theta)}$ is defined in (\ref{fpsinn}). Then, the CDF of type-I central angle distribution can be written as follows,
\begin{equation}\label{appD-1}
\begin{split}
    &F_{\theta_c^{(1)}} \left(\Xi\right) = \int_0^{2\pi} \int_0^{\Xi} f_{\psi_{nn}, \phi_{nn}}(\theta_{r2n}) R_s^2 \sin\theta \mathrm{d}\theta \mathrm{d}\phi \\
    &= \int_0^{2\pi} \int_0^{\Xi} f_{\psi_{nn}, \phi_{nn}}\left( 2\arcsin \frac{d(\theta,\phi;\frac{\Theta}{N_l},0)}{2R_s} \right) R_s^2 \sin\theta \mathrm{d}\theta \mathrm{d}\phi
\end{split}
\end{equation}
where $f_{\psi_{nn}}(\theta_{r2n})$ and $d(\theta,\phi;\frac{\Theta}{N_l},0)$ are defined in (\ref{fpsinn}) and (\ref{d_12}) respectively, $\theta_{r2n}$ is the central angle between reference point $\left( R_s,0,0 \right)$ and nearest neighbor of $\left( R_s,\frac{\Theta}{N_l},0 \right)$, $R_s^2 \sin\theta$ is the area element in spherical coordinates with a fixed radius $R_s$.

\par
When the central angle is smaller than $\Xi$, the Euclidean distance between $\left( R_s,0,0 \right)$ and nearest neighbor is $2R_s\sin\frac{\Xi}{2}$. Refer to the concept of multiplicative distance scale factor, in the derivation of type-II central angle distribution, the distance between two neighbors of reference points distance goes up by $\alpha^{(1)}\left(\frac{\Theta}{N_l}\right)$ times,
\begin{sequation}
\begin{split}
    F_{\theta_c^{(2)}} \left(\Xi\right) & = \mathbb{P}\left[ \theta_c^{(2)} < \Xi \right] = \mathbb{P}\left[ 2R_s\sin\frac{\theta_c^{(2)}}{2} < 2R_s\sin\frac{\Xi}{2} \right] = \mathbb{P}\left[ \alpha^{(1)}\left(\frac{\Theta}{N_l}\right) 2R_s\sin\frac{\theta_c^{(1)}}{2} < 2R_s\sin\frac{\Xi}{2} \right] \\
    & = \mathbb{P}\left[ \theta_c^{(1)} < 2 \arcsin \frac{\sin\frac{\Xi}{2}}{\alpha^{(1)}\left(\frac{\Theta}{N_l}\right)}  \right] = F_{\theta_c^{(1)}} \left(2 \arcsin \frac{\sin(\Xi/2)}{\alpha^{(1)}\left( \Theta/N_l \right)}\right).
\end{split}
\end{sequation}

Finally, the probability of the route is available can be given according to its definition,
\begin{equation}
\begin{split}
    P^A = \mathbb{P}\left[ 2R_s \sin \frac{\theta_c^{(1)}}{2}<2\sqrt{R_s^2-R_{\oplus}^2} \right]^2 
    \times \mathbb{P}\left[ 2R_s \sin \frac{\theta_c^{(2)}}{2}<2\sqrt{R_s^2-R_{\oplus}^2} \right]^{N_l-2}. 
\end{split}
\end{equation}

\section{Proof of Theorem~\ref{theorem3}}\label{app:theorem3}
According to the definition in (\ref{deflatency}), the average routing transmission latency can be obtained by taking the expectation of the random variables,
\begin{sequation}
\begin{split}
    & \overline{T}_{\rm tx} = \sum_{i=1}^{N_l} \mathbb{E}_{l_i}\left[ \mathbb{E}_{W}\left[ \frac{\varpi}{B}  \left(\log_2 \left(1+\rho^t \, G \left( \frac{\lambda}{2\pi l_i} \right)^2 \sigma^{-2} W \right)\right)^{-1} \right]  \right],
\end{split}
\end{sequation}
where the random variable $l_i$ is the distance of the $i^{th}$ hop. The procedure is relatively straightforward, therefore omitted here.

When communication is ultra-reliable (${\rm SNR} \gg 1$), we have
\begin{sequation}
\begin{split}
    & \mathbb{E}_W \left[ \frac{\varpi}{B} \left(\log_2 \left(1+\rho^t \, G \left( \frac{\lambda}{2\pi l_i} \right)^2 \sigma^{-2} W \right)\right)^{-1} \right] \\
    & \overset{(a)}{\gtrsim}  \frac{\varpi}{B} \left(\log_2 \left(1 + \rho^t \, G \left( \frac{\lambda}{2\pi l_i} \right)^2 \sigma^{-2} \mathbb{E}_W \left[W\right] \right)\right)^{-1} \approx  T_{{\rm tx},1}^*(l_i), 
\end{split}
\end{sequation}
where step (a) follows Jensen's inequality, under the condition that $f_W(w)=\left(\log_2 \left( 1+cw \right)\right)^{-1}$ is a convex function of $w$. $c=\rho^t \, G \left( \frac{\lambda}{2\pi l_i} \right)^2 \sigma^{-2}$ is the SNR without small-scale fading. When $c > {\rm SNR} \gg 1$, step (a) is the major source of the approximation error. The Jensen gap decreases with $c$, because
\begin{equation}
\begin{split}
    \frac{f_W(w+\Delta w)-f_W(w)}{w+\Delta w - w} \approx \frac{\log_2 w - \log_2 \left( w+\Delta w \right)}{\Delta w \, \log_2\left(cw+c\Delta w\right) \, \log_2\left(cw\right)}.
\end{split}
\end{equation}



\bibliographystyle{IEEEtran}
\bibliography{references}

\end{document}